\documentclass[journal]{IEEEtran}
\usepackage{amssymb}
\usepackage{amsfonts}
\usepackage{amsfonts,subfigure,multicol,color,verbatim,
graphicx,cite,epsfig,amssymb,amsmath,cases,bm,algorithm,
algorithmic,xcolor,multirow,array,booktabs}
\usepackage{diagbox} 
\usepackage{rotating}

\begin{document}

\title{Application of Machine Learning in Wireless Networks: Key Techniques and Open Issues}

\author{Yaohua~Sun, Mugen~Peng, \IEEEmembership{Senior Member, IEEE}, Yangcheng Zhou, Yuzhe Huang, and Shiwen~Mao, \IEEEmembership{Fellow, IEEE}
\thanks{Yaohua~Sun (e-mail: sunyaohua@bupt.edu.cn), Yangcheng Zhou (e-mail: bboyjinjosun@sina.com), and Yuzhe Huang (e-mail: 467292619@qq.com) are with the Key Laboratory of Universal Wireless Communications (Ministry of Education), Beijing University of Posts and Telecommunications, Beijing, China.
Mugen~Peng (e-mail: pmg@bupt.edu.cn) is with the State Key Laboratory of Networking and Switching Technology (SKL-NST),
Beijing University of Posts and Telecommunications, Beijing, China.
Shiwen Mao (smao@ieee.org) is with the Department of Electrical and Computer Engineering, Auburn University, Auburn, AL 36849-5201, USA. {\bf (Corresponding author: Shiwen Mao)}}
\thanks{This work is supported in part by XXXX, and the US National Science Foundation under grant CNS-1702957.}}

\maketitle

\begin{abstract}
As a key technique for enabling artificial intelligence, machine learning (ML) is capable of solving complex problems without explicit programming.
Motivated by its successful applications to many practical tasks like image recognition, both industry and the research community have
advocated the applications of ML in wireless communication.
This paper comprehensively surveys the recent advances of the applications
of ML in wireless communication, which are classified as: resource management in the MAC layer, networking and mobility management in the network layer, and localization in the application layer. The applications in resource management further include power control, spectrum management, backhaul management, cache management, beamformer design
and computation resource management,
while
ML based networking focuses on the applications in clustering, base station switching control, user association and routing.
Moreover, literatures in each aspect is organized according to the adopted ML techniques.
In addition, several conditions for applying ML to wireless communication are identified to help readers decide
whether to use ML and which kind of ML techniques to use, and traditional approaches are also summarized together with their performance comparison
with ML based approaches, based on which the motivations of surveyed literatures to adopt ML are clarified.
Given the extensiveness of the research area, challenges and unresolved issues are presented to facilitate future studies,
where ML based network slicing, infrastructure update to support ML based paradigms, open data sets and platforms for researchers, theoretical guidance for ML implementation and so on are discussed.
\end{abstract}



\section{Introduction}

Since the rollout of the first generation wireless communication system,
wireless technology has been continuously evolving from supporting
basic coverage to satisfying more advanced needs. In particular, the fifth generation (5G)
mobile communication system is expected to achieve a considerable increase in data rates, coverage
and the number of connected devices with latency and
energy consumption significantly reduced\cite{5G}. Moreover,
5G is also expected to provide more accurate localization, especially in an indoor environment \cite{position}.

These goals can be potentially met by enhancing the system from different aspects.
For example, computing and caching resources can be deployed at the network edge to fulfill the demands
for low latency and reduce energy consumption \cite{mecsurvey,cacheair}, and the cloud computing based BBU pool can provide
high data rates with the use of large-scale collaborative signal processing among BSs and can save much
energy via statistical multiplexing \cite{sunsurvey}. Furthermore,
the co-existence of heterogenous nodes,
including macro BSs (MBSs), small base stations (SBSs) and user equipments (UEs) with device-to-device (D2D) capability,
can boost the throughput and simultaneously guarantee seamless coverage \cite{d2dsurvey}.
However, the involvement of computing resource, cache resource and heterogenous nodes cannot alone satisfy the stringent requirements of 5G.
The algorithmic design enhancement for resource management, networking, mobility management and localization is essential as well.
Faced with the characteristics of 5G, current resource management, networking, mobility management and localization algorithms
expose several limitations.

First, with the proliferation of smart phones, the expansion of network scale and the diversification of services in the 5G era,
the amount of data, related to applications, users and networks, will experience an explosive growth,
which can contribute to an improved system performance if properly utilized \cite{data}.
However, many of the existing algorithms are incapable of processing and/or utilizing the data,
meaning that much valuable information or patterns are wasted.
Second,
to adapt to the dynamic network environment, algorithms like RRM algorithms are often fast but
heuristic. Since the resulting system performance can be far from optimal, these algorithms can hardly
meet the performance requirements of 5G. To obtain better performance,
research has
been done based on optimization theory to develop more effective algorithms to reach optimal or
suboptimal solutions. However, many studies assume a static network environment.
Considering that 5G networks will be more complex, hence leading to more complex mathematical formulations,
the developed algorithms can possess high complexity. Thus, these algorithms will be inapplicable in the real dynamic
network, due to their long decision-making time.
Third, given the large number of nodes in future 5G networks, traditional centralized algorithms for network management
can be infeasible due to the high computing burden and high cost to collect global information.
Therefore, it is preferred to enable network nodes to autonomously make decisions based on local observations.

As an important enabling technology for artificial intelligence, machine learning has been successfully applied in many areas,
including computer vision, medical diagnosis, search engines and speech recognition \cite{machine}.
Machine learning is a field of study that gives computers the ability to learn without being explicitly programmed.
Machine learning techniques can be generally classified as supervised learning, unsupervised learning and reinforcement learning.
In supervised learning, the aim of the learning agent is to learn a general rule mapping inputs to outputs with
example inputs and their desired outputs provided, which constitute the labeled data set. In unsupervised learning,
no labeled data is needed, and the agent tries to find some structures from its input. While in reinforcement learning,
the agent continuously interacts with an environment and tries to generate a good policy according to the immediate reward/cost
fed back by the environment.
In recent years, the development of fast and massively parallel graphical processing
units and the significant growth of data have contributed to the progress in deep learning,
which can achieve more powerful representation capabilities.
For machine learning, it has the following advantages to overcome the drawbacks of traditional resource management, networking,
mobility management and localization algorithms.

The first advantage is that machine learning has the ability to learn useful information from input data, which can help improve network performance.
For example, convolutional neural networks and recurrent neural networks can extract spatial features and sequential features from
time-varying Received Signal Strength Indicator (RSSI), which can mitigate the ping-pong effect in mobility management \cite{1},
and more accurate indoor localization for a three-dimensional space can be achieved by using an auto-encoder to extract robust fingerprint patterns from noisy RSSI measurements \cite{3d}.
Second, machine learning based resource management, networking and mobility management algorithms can well adapt to the dynamic environment.
For instance, by using the deep neural network proven to be an universal function approximator, traditional high complexity algorithms can be
closely approximated, and similar performance can be achieved but with much lower complexity \cite{nnbeam}, which makes it possible to quickly response to environmental changes.
In addition, reinforcement learning can achieve fast network control based on learned policies \cite{bs3}.
Third, machine learning helps to realize the goal of network self-organization. For example,
using multi-agent reinforcement learning, each node in the network can self-optimize its transmission power, subchannel allocation and so on.
At last, by involving transfer learning, machine learning has the ability to quickly solve a new problem.
It is known that there exist some temporal and spatial relevancies in wireless systems such as
traffic loads between neighboring regions \cite{bs3}. Hence, it is possible to transfer the knowledge acquired in one task to another relevant task, which
can speed up the learning process for the new task. However, in traditional algorithm design, such prior knowledge is often not utilized.
In Section VIII, a more comprehensive discussion about the motivations to adopt machine learning rather than traditional approaches is made.

Driven by the recent development of the applications of machine learning in wireless networks,
some efforts have been made to survey related research and
provide useful guidelines.
In \cite{jiang}, the basics of some machine learning algorithms along with their applications in future wireless networks are introduced, such as
Q-learning for the resource allocation and interference coordination in downlink femtocell networks and Bayesian learning for channel parameter estimation
in a massive, multiple-input-multiple-output network.
While in \cite{chuangan}, the applications of machine learning in wireless sensor networks (WSNs) are discussed, and the advantages and disadvantages
of each algorithm are evaluated with guidance provided for WSN designers.
In \cite{wulian},
different learning techniques, which are suitable for
Internet of Things (IoT), are presented, taking into account the unique characteristics of IoT, including resource constraints and strict quality-of-service requirements,
and studies on learning for IoT are also reviewed in \cite{iotreview}.
The applications of machine learning in cognitive radio (CR) environments are investigated in \cite{cognitive} and \cite{cog1}.
Specifically, authors in \cite{cognitive} classify those applications into decision-making tasks and
classification tasks, while authors in \cite{cog1} mainly concentrate on model-free strategic learning.
Moreover, authors in \cite{access} and \cite{son} focus on the potentials of machine learning in enabling the self-organization of cellular networks with the
perspectives of self-configuration, self-healing and self-optimization.
To achieve high energy efficiency in wireless networks, related promising approaches based on big data
are summarized in \cite{energy}.
In \cite{chensurvey}, a comprehensive tutorial on the applications of neural networks (NNs) is provided, which presents the
basic architectures and training procedures of different types of NNs, and several typical application scenarios
are identified.
In \cite{arxiv1} and \cite{arxiv2}, the applications of deep learning in the physical layer are summarized.
Specifically,
authors in \cite{arxiv1} see the whole communication system as an auto-encoder, whose task is to learn compressed representations of user messages that
are robust to channel impairments. A similar idea is presented in \cite{arxiv2}, where a broader area is surveyed including
modulation recognition, channel decoding and signal detection.
Also focusing on deep learning, literatures \cite{intelligent,networking,S-23} pay more attention to upper layers,
and the surveyed applications include channel resource allocation, routing, scheduling, and so on.

Although significant progress has been achieved toward surveying the applications of machine learning in wireless networks,
there still exist some limitations in current works.
More concretely, literatures \cite{jiang,cognitive,access,son,energy} are seldom related to deep learning, deep reinforcement learning and transfer learning,
and the content in \cite{arxiv1,arxiv2} only focuses on the physical layer.
In addition, only a specific network scenario is covered in \cite{chuangan,wulian,iotreview,cognitive,cog1},
while literatures \cite{chensurvey,intelligent,networking,S-23} just pay attention to NNs or deep learning.
Considering these shortcomings and the ongoing research activities, a more comprehensive
survey framework, which incorporates the recent achievements in the applications of diverse machine learning methods in different layers and network scenarios,
seems timely and significant.
Specifically,
this paper surveys state-of-the-art applications of various machine learning approaches from the MAC layer up to the application layer,
covering resource management, networking, mobility
management and localization.
The principles of different learning techniques are introduced, and useful guidelines are provided.
To facilitate future applications of machine learning,
challenges and open issues are identified.
Overall, this survey aims to fill the gaps found in the previous papers \cite{jiang,chuangan,wulian,iotreview,cognitive,cog1,access,son,energy,chensurvey,arxiv1,arxiv2,intelligent,networking,S-23}, and our contributions are threefold:
\begin{enumerate}
 \item Popular machine learning techniques utilized in wireless networks are comprehensively summarized including their basic principles and general applications, which are classified into supervised learning, unsupervised learning, reinforcement learning, (deep) NNs and transfer learning. Note that (deep) NNs and transfer learning are separately highlighted because of their increasing importance to wireless communication systems.
 \item A comprehensive survey of the literatures applying machine learning to resource management, networking, mobility management and localization is presented, covering all the layers except the physical layer that has been thoroughly investigated. Specifically, the applications in resource management are further divided into power control, spectrum management, beamformer design, backhaul management, cache management and computation resource management, and the applications in networking are divided into user association, BS switching control, routing and clustering. Moreover, surveyed papers in each application area are organized according to their adopted machine learning techniques, and the majority of the network scenarios that will emerge in the 5G era are included such as vehicular networks, small cell networks and cloud radio access networks.
 \item Several conditions for applying machine learning in wireless networks are identified, in terms of problem types, training data availability, time cost, and so on. In addition, the traditional approaches, taken as baselines in surveyed literatures, are summarized, along with performance comparison with machine learning based approaches, based on which motivations to adopt machine learning are clarified.
     Furthermore, future challenges and unsolved issues related to the applications of machine learning in wireless networks are discussed with regard to
 machine learning based network slicing, standard data sets for research, theoretical guidance for implementation, and so on.
\end{enumerate}

The remainder of this paper is organized as follows: Section II
introduces the popular machine learning techniques utilized in wireless networks,
together with a brief summarization of their applications.
The applications of machine learning in resource management are summarized in Section III.
In Section IV, the applications of machine learning in networking are surveyed.
Section V and VI summarize recent advances in machine learning based mobility management and localization, respectively.
Several conditions for applying machine learning are presented in Section VII to help readers decide
whether to adopt machine learning and which kind of learning techniques to use.
In Section VIII, traditional approaches and their performance compared with machine learning based approaches are summarized,
and then motivations in surveyed literatures to use machine learning are elaborated.
Challenges and open issues are identified in Section IX, followed by the conclusion in Section X.
For convenience, all abbreviations are listed in Table I.

\begin{table*}\label{table2}
\center \caption{SUMMARY OF ABBREVIATIONS}
\begin{tabular}{l l | l l}
 \toprule
3GPP & $3^{rd}$ generation partnership project  & NN & neural network\\
5G & $5^{th}$ generation  & OFDMA & orthogonal frequency division multiple access \\
AP & access point & OISVM & online independent support vector machine \\
BBU & baseband unit & OPEX & operating expense \\
BS & base station  & OR & outage ratio \\
CBR & call blocking ratio & ORLA & online reinforcement learning approach \\
CDR & call dropping ratio & OSPF & open shortest path first \\
CF & collaborative filtering & PBS & pico base station \\
CNN & convolutional neural network & PCA & principal components analysis \\
CR & cognitive radio  & PU & primary user \\
CRE & cell range extension & QoE & quality of experience \\
CRN & cognitive radio network & QoS & quality of service \\
C-RAN & cloud radio access network & RAN & radio access network  \\
CSI & channel state information & RB & resource block  \\
D2D & device-to-device  & ReLU & rectified linear unit \\
DRL & deep reinforcement learning & RF & radio frequency \\
DNN & dense neural network & RL & reinforcement learning \\
EE & energy efficiency  & RNN & recurrent neural network \\
ELM & extreme learning machine & RP & reference point \\
ESN & echo state network & RRH & remote radio head \\
FBS & femto base station & RRM & radio resource management\\
FLC & fuzzy logic controller & RSRP & reference signal receiving power  \\
FS-KNN & feature scaling based k nearest neighbors & RSRQ & reference signal receiving quality  \\
GD & gradient descent & RSS & received signal strength \\
GPS & global positioning system & RSSI & received signal strength indicator \\
GPU & graphics processing unit & RVM & relevance vector machine \\
Hys & hysteresis margin & SBS & small base stations  \\
IA & interference alignment & SE & spectral efficiency \\
ICIC & inter-cell interference coordination & SINR & signal-to-interference-plus-noise ratio \\
IOPSS & inter operator proximal spectrum sharing & SNR & signal-to-noise ratio \\
IoT & internet of things & SOM & self-organizing map \\
KNN & k nearest neighbors & SON & self-organizing network \\
KPCA & kernel principal components analysis & SU & secondary user \\
KPI & key performance indicators & SVM & support vector machine \\
LOS & line-of-sight & TDMA & time division multiple access \\
LSTM & long short-term memory & TDOA & time difference of arrival \\
LTE & long term evolution & TOA & time of arrival \\
LTE-U & long term evolution-unlicensed & TTT & time-to-trigger \\
MAB & multi-armed bandit & TXP & transmit power \\
MAC & medium access control & UE & user equipment \\
MBS & macro base station & UAV & unmanned aerial vehicle  \\
MEC & mobile edge computing & UWB & ultra-wide bandwidth  \\
ML & machine learning & WLAN & wireless local area network \\
MUE & macrocell user equipment & WMMSE & weighted minimum mean square error \\
NE & Nash equilibrium & WSN & wireless sensor network \\
NLOS & non-line-of-sight & &\\
 \bottomrule
\end{tabular}
\end{table*}


\section{Machine Learning Preliminaries}

In this section, various machine learning techniques
employed in the studies surveyed in this paper are introduced, including
supervised learning, unsupervised learning, reinforcement learning, (deep) NNs and transfer learning.

\subsection{Supervised Learning}

Supervised learning is a machine learning task that aims to learn a mapping function from the input to the output, given
a labeled data set. Specifically, supervised learning can be further divided into regression and classification based on the continuity of the output.
In surveyed works, the following supervised learning techniques are adopted.

\subsubsection{Support Vector Machine}


The basic support vector machine (SVM) model is a linear classifier, which aims to separate data points that are $p$-dimensional vectors using a $p-1$ hyperplane.
The best hyperplane is the one that leads to the largest separation or margin between the two given classes, and
is called the maximum-margin hyperplane. However, the data set is often not linearly separable in the original space. In this case,
the original space can be mapped to a much higher dimensional space by involving kernel functions, such as polynomial or Gaussian kernels,
which results in a non-linear classifier. More details about SVM can be found in \cite{svm}.


\subsubsection{K Nearest Neighbors}


K Nearest Neighbors (KNN) is a non parametric lazy learning algorithm for classification and regression, where no assumption on the data distribution
is needed. Taking the classification task as an example, the basic principle of KNN is to decide the class of a test point in the form of a feature vector
based on the majority voting of its K nearest neighbors.
Considering that training data belonging to very frequent classes
can dominate the prediction of test data, a weighted method can be adopted by involving a weight for each neighbor that is proportional to the inverse of its distance to the test point. In addition, one of the keys to applying KNN is the tradeoff of the parameter $K$, and the value selection can be referred to \cite{knn}.


\subsection{Unsupervised Learning}

Unsupervised learning is a machine learning task that aims to learn a function to describe a hidden structure from unlabeled data.
In surveyed works, the following unsupervised learning techniques are utilized.

\subsubsection{K-Means Clustering Algorithm}

\begin{figure}[!t]
\centering
\includegraphics[scale=0.5]{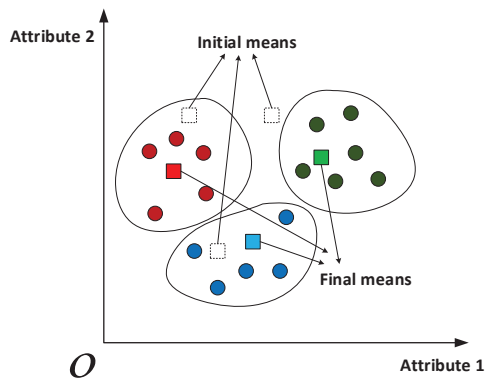}
\caption{An illustration of the K-means model.}\label{kmeans}
\end{figure}

In K-means clustering, the aim is to partition $n$ data points into $K$ clusters, and each data point belongs to the cluster with the nearest mean.
The most common version of K-means algorithm is based on iterative refinement. At the beginning, $K$ means are randomly initialized. Then, in each iteration,
each data point is assigned to exactly one cluster, whose mean has the least Euclidean distance to the data point, and
the mean of each cluster is updated. The algorithm continuously iterates until the members in each cluster do not change.
The basic principle is illustrated in Fig. \ref{kmeans}.

\subsubsection{Principal Component Analysis}

Principal component analysis (PCA) is a
dimension-reduction tool that can be used to
reduce a large set of variables to a small set
that still contains most of the information in
the large set and is mathematically defined as an orthogonal linear transformation that transforms the data to a new coordinate system.
More specifications of PCA can be referred to \cite{pcaprin}.

\subsection{Reinforcement Learning}

In reinforcement learning (RL), the agent aims to optimize a long term objective by interacting with the environment based on a trial and error process.
Specifically, the following reinforcement learning algorithms are applied in surveyed studies.

\subsubsection{Q-Learning}


One of the most commonly adopted reinforcement learning algorithms is Q-learning. Specifically,
the RL agent interacts with the environment to learn the Q values,
based on which the agent takes an action. The Q value is defined as the discounted accumulative reward, starting at a tuple of a state and an action and then
following a certain policy.
Once the Q values are learned after a sufficient amount of time, the agent can make a quick decision under the current state by taking the action with
the largest Q value. More details about Q learning can be referred to \cite{q}.
In addition, to handle continuous state spaces, fuzzy Q learning can be used \cite{fuzzyq}.


\subsubsection{Multi-Armed Bandit Learning}


In a multi-armed bandit (MAB) model with a single agent, the agent sequentially takes an action and then receives a random reward generated by a corresponding distribution, aiming
at maximizing an aggregate reward. In this model, there exists a tradeoff between taking the current, best action (exploitation) and gathering information to
achieve a larger reward in the future (exploration). While in the MAB model with multiple agents, the reward an agent receives after playing an action
is not only dependent on this action but also on the agents taking the same action. In this case, the model is expected to achieve some steady states or equilibrium \cite{ha}.
More details about MAB can be referred to \cite{MAB}.


\subsubsection{Actor-Critic Learning}

\begin{figure}[!t]
\centering
\includegraphics[scale=0.5]{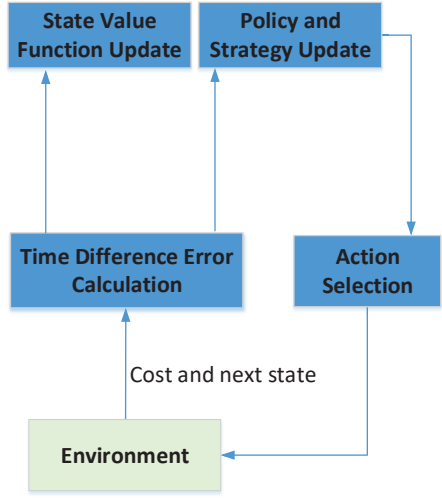}
\caption{The process of actor-critic learning.}\label{actor}
\end{figure}

The actor-critic learning algorithm is composed of an actor, a critic and an environment with which the actor interacts.
In this algorithm, the actor first selects an action according to the current strategy and receives an immediate cost. Then, the critic updates
the state value function based on a time difference error, and next, the actor will update the policy.
As for the strategy, it can be updated based on learned policy using Boltzmann distribution.
When each action is revisited infinitely for each state, the algorithm will converge to the optimal state values \cite{actorconver}.
The process of actor-critic learning is shown in Fig. \ref{actor}.

\subsubsection{Joint Utility and Strategy Estimation Based Learning}

\begin{figure}[!t]
\centering
\includegraphics[scale=0.4]{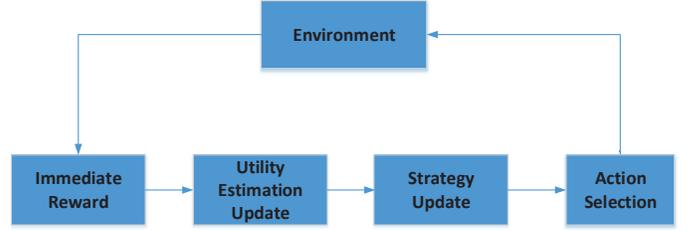}
\caption{The process of joint utility and strategy estimation based learning.}\label{jumodel}
\end{figure}

In this algorithm shown in Fig. \ref{jumodel}, each agent holds an estimation of the expected utility, whose update is based on the immediate reward, and the probability
to select each action, named as strategy, is updated in the same iteration based on the utility estimation \cite{ju}.
The main benefit of this algorithm lies in that it is fully distributed when the reward can be directly calculated locally, as, for example,
the data rate between a transmitter and its paired receiver.
Based on this algorithm, one can further estimate the regret of each action based on utility estimations and the received immediate reward, and then update strategy using regret estimations. In surveyed works, this algorithm is often connected with some equilibrium concepts in game theory like Logit equilibrium and coarse correlated
equilibrium.

\subsubsection{Deep Reinforcement Learning}

\begin{figure}[!t]
\centering
\includegraphics[width=3.4in]{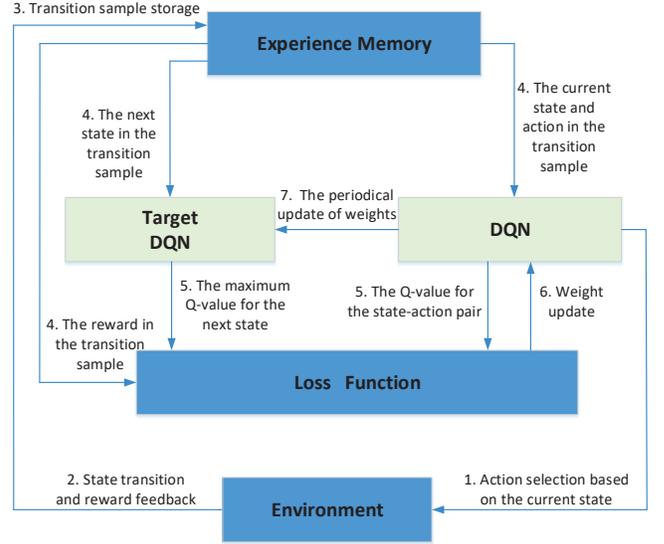}
\caption{The main components and process of the basic DRL.}\label{drl}
\end{figure}

In \cite{dqn}, authors propose to use a deep NN, called deep Q network (DQN), to approximate optimal Q values, which allows the agent to
learn from the high-dimensional sensory data directly, and reinforcement learning based on DQN is known as deep reinforcement learning (DRL).
Specifically, state transition samples generated by interacting with the environment are stored in the replay memory
and sampled to train the DQN,
and a target DQN is adopted to generate target values, which both help stabilize the training procedure of DRL.
Recently, some enhancements to DRL have come out, and readers can refer to the literature \cite{qsurvey} for more details.
The main components and working process of the basic DRL are shown in Fig. \ref{drl}.

\subsection{(Deep) Neural Network}



\subsubsection{Dense Neural Network}

\begin{figure}[!t]
\centering
\includegraphics[scale=0.5]{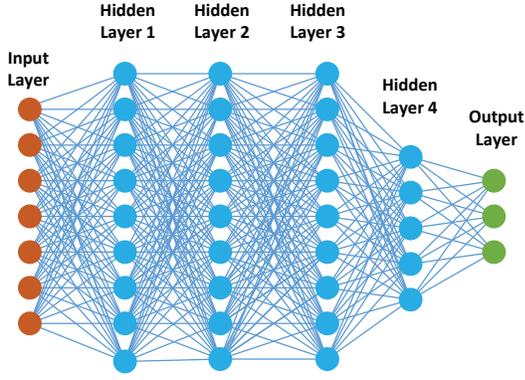}
\caption{The architecture of a DNN.}\label{dense}
\end{figure}

As shown in Fig. \ref{dense}, the basic component of a dense neural network (DNN) is a neuron corresponding with weights for the input and an activation function for
the output. Common activation functions include tanh, Relu, and so on.
The input is transformed through the network layer by layer, and there is no direct connection between two non-consecutive layers.
To optimize network parameters, backpropagation together with various GD methods can be employed, which include
Momentum, Adam, and so on \cite{gdopt}.

\subsubsection{Recurrent Neural Network}

\begin{figure}[!t]
\centering
\includegraphics[scale=0.4]{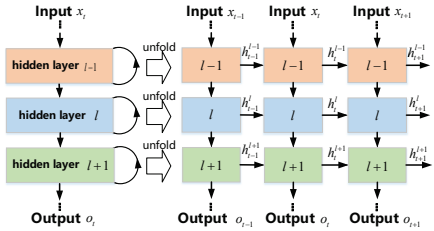}
\caption{The architecture of an RNN.}\label{rnn}
\end{figure}

As demonstrated in Fig. \ref{rnn},
there exist connections in the same hidden layer in the recurrent neural network (RNN) architecture.
After unfolding the architecture along the time line,
it can be clearly seen that the output at a certain time step is
dependent on both the current input and former inputs, hence RNN
is capable of remembering.
RNNs include echo state networks, LSTM, and so on.

\subsubsection{Convolutional Neural Network}

\begin{figure}[!t]
\centering
\includegraphics[scale=0.5]{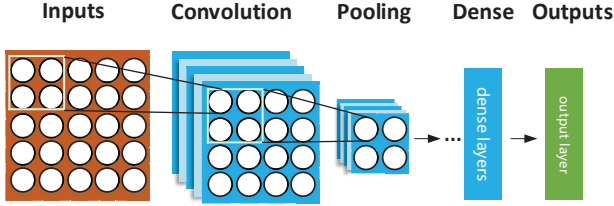}
\caption{The architecture of a CNN.}\label{cnn}
\end{figure}

As shown in Fig. \ref{cnn},
two main components of convolutional neural networks (CNNs) are layers for convolutional operations and pooling operations like maximum pooling.
By stacking convolutional layers and pooling layers alternately, the CNN can learn rather complex models based on progressive levels of abstraction.
Different from dense layers in DNNs that learn global patterns of the input, convolutional layers can learn local patterns. Meanwhile,
CNNs can learn spatial hierarchies of patterns.

\subsubsection{Auto-encoder}

\begin{figure}[!t]
\centering
\includegraphics[scale=0.5]{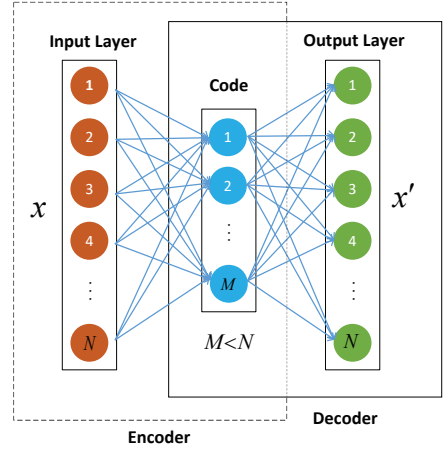}
\caption{The structure of an autoencoder.}\label{auto}
\end{figure}

An auto-encoder is an unsupervised neural network with the target output the same as the input,
and it has different
variations like denoising auto-encoder and sparse auto-encoder. With the limited number of
neurons, an auto-encoder can learn a compressed but robust representation of the input in order to construct the input at the output.
Generally, after an auto-encoder is trained, the decoder is removed with the encoder kept
as the feature extractor. The structure of an auto-encoder is demonstrated in Fig. \ref{auto}.
Note that auto-encoders have some disadvantages. First, a pre-condition for a trained auto-encoder to work well is that the distribution of
the input data had better be identical to that of training data. Second, its working mechanism is often seen as a black box that can not be clearly explained.
Third, the hyper-parameter setting, which is a complex task, has a great influence on its performance, such as the number of neurons in the hidden layer.

\subsubsection{Extreme Learning Machine}

The extreme learning machine (ELM) is a
feed-forward NN with a single or multiple hidden-layers whose
parameters are randomly generated using a
distribution, while the weights between the hidden layer and output layer are computed
by minimizing the error between the computed output value
and the true output value.
More specifications on ELM can be referred to \cite{elmbasic}.


\subsection{Transfer Learning}

Transfer learning is a concept with the goal of utilizing the knowledge
from a specific domain to assist learning in a target domain.
By applying transfer learning, which can avoid learning from scratch,
the new learning process can be sped up.
In deep learning and reinforcement learning, knowledge can be represented by weights and Q values, respectively.
Specifically, when deep learning is adopted for image recognition, one can use the weights that have been well trained for another
image recognition task as the initial weights, which can help achieve a satisfactory performance with a small training set.
For reinforcement learning, Q values learned by an agent in a former environment can be involved in the Q value update in a new
but similar environment to make a wiser decision at the initial stage of learning. Specifications on integrating transfer learning with
reinforcement learning can be referred to \cite{qianghuaqianyi}.
However, when transfer learning is utilized, the negative impact of former knowledge on the performance should be carefully handled,
since there still exist some differences between tasks or environments.

\subsection{Some Implicit Assumptions}

For SVM and neural network based supervised learning methods adopted in surveyed works,
they actually aim at learning a mapping rule from the input to the output based on training data.
Hence, to apply the trained supervised learning model to make effective predictions, an implicit assumption in surveyed literatures
is that the learned mapping rule still works in future environments.
In addition,
there are also some implicit assumptions in surveyed literatures applying reinforcement learning.
For reinforcement learning with a single agent, the dynamics in the studied communication system,
such as cache state transition probabilities, should be unchanged.
This is intuitive, because the learning agent will face a new environment once the dynamics change,
which invalidates the learned policy.
As for reinforcement learning with multiple agents whose policies or strategies are coupled,
an assumption to use it is that
the set of agents should be kept fixed, since the variation of the agent set will also
make the environment different from the perspective of each agent.

At last, to intuitively show the applications of each machine learning method, Table II is drawn with all the surveyed works
listed.

\begin{sidewaystable*}
\caption{The applications of machine learning methods surveyed in this paper}
\scriptsize
\centering
\begin{tabular}{|p{4.1cm}|p{1cm}|p{1cm}|p{1cm}|p{1cm}|p{1cm}|p{0.1cm}|p{1cm}|p{1cm}|p{1cm}|p{1cm}|p{1.2 cm}|p{1.2 cm}|}
\hline
\multicolumn{1}{|c|}{\multirow{2}{4.1cm}{\diagbox{Learning Methods}{Applications}}} & \multicolumn{6}{c|}{Resource Management} & \multicolumn{4}{c|}{Networking} &
\multicolumn{1}{c|}{\multirow{2}{1.2cm}{Mobility Management}} & \multicolumn{1}{c|}{\multirow{2}{1.2cm}{Localization}}\\
\cline{2-11}
& \multicolumn{1}{c|}{Power} & \multicolumn{1}{c|}{Spectrum} & \multicolumn{1}{c|}{Backhaul} &  \multicolumn{1}{c|}{Cache} & \multicolumn{1}{c|}{Beamformer} & \multicolumn{1}{c|}{Computation Resource} &\multicolumn{1}{c|}{User Association} & \multicolumn{1}{c|}{BS Switching} & \multicolumn{1}{c|}{Routing} & \multicolumn{1}{c|}{Clustering} & & \\
\hline
Support Vector Machine &[47] [48]& & &[75]& & & & & & & & [119] [121] \\
\hline
K Nearest Neighbors & & & & &[80]& & & & & & & [116] [117] [118] \\
\hline
K-Means Clustering & & & & & & & &[96] [97]& &[106] [96] & & \\
\hline
Principal Component Analysis & & & & & & & & & & & & [125] [126]\\
\hline
Actor-Critic Learning & & & & & & & &[88] [12] [94] [95]& & & & \\
\hline
Q-Learning &[42] [43] [44] [46] [51]&[52] [56] &[63] [60] [59] [65]&[69] & & &[82] [83] & [90] [89] [98] [93] [92] [97]& [99] [100] [102] [101]& &[108] [109] [110] [111] & \\
\hline
Multi-Armed Bandit Learning & &[53] & & & & &[33] & & & & & \\
\hline
Joint Utility and Strategy Estimation Based Learning &[45] &[55] &[64] [62] [61]&[68] & & & & & & & & \\
\hline
Deep Reinforcement Learning & & & &[70] [71]  & & [79]&[85] &[91] & & [107]& [10] [112] & \\
\hline
Auto-encoder &[50] & & & & & & & & & & & [127] [124] [128] [129] [11]\\
\hline
Extreme-Learning Machine & & & &[72] & & & & & & & & [122]\\
\hline
Dense Neural Network &[9]& & & & & & & & [103] &[105] & & [123]\\
\hline
Convolutional Neural Network &[49]& & & [75] & & & & &[104] & & & \\
\hline
Recurrent Neural Network & & [57] [58]& &[73] [74] & & & & & & & & \\
\hline
Transfer Learning &[51]& & &[76] [77] [78]& & & & [12] [94] [95] [97]& & & & \\
\hline
Collaborative Filtering & & &  & & & &[84] & & & & & \\
\hline
Other Reinforcement Learning Techniques & & [54]&  & & & &[81] & & & & & \\
\hline
Other Supervised/Unsupervied Techniques & & &  & & & & & & & & [113] [114] [115]&[120] \\
\hline
\end{tabular}
\end{sidewaystable*}

\section{Machine Learning Based Resource Management}

In wireless networks, resource management aims to achieve proper utilization of limited physical resources to
meet various traffic demands and improve system performance.
Academic resource management methods are often designed for static networks and highly dependent on formulated mathematical problems.
However, the states of practical wireless networks are dynamic, which will lead to the frequent re-execution of algorithms that can possess high complexity,
and meanwhile, the ideal assumptions facilitating the formulation of a tractable mathematical problem can result in large performance loss
when algorithms are applied to real situations.
In addition, traditional resource management can be enhanced by extracting useful information related to users and networks,
and distributed resource management schemes are preferred when the number of nodes is large.

Faced with the above issues, machine learning techniques including model-free reinforcement learning and NNs can be employed.
Specifically, reinforcement learning can learn a good resource management policy based on only the reward/cost
fed back by the environment, and
quick decisions can be made for a dynamic network once a policy is learned.
In addition,
owing to the superior approximation capabilities of deep NNs, some high complexity resource management algorithms can be approximated, and
similar network performance can be achieved but with much lower complexity.
Moreover, NNs can be utilized to learn the content popularity, which helps fully make use of limited cache resource,
and distributed Q-learning can endow each node with autonomous decision capability for resource allocation.
In the following, the applications of machine learning in power control, spectrum management, backhaul management, beamformer design, computation resource management and cache management will be introduced.

\subsection{Machine Learning Based Power Control}

In the spectrum sharing scenario, effective power control can reduce inter-user interference, and hence increase system throughput.
In the following, reinforcement, supervised, and transfer learning-based power control are elaborated.

\subsubsection{Reinforcement Learning Based Approaches}


In \cite{24}, authors focus on inter-cell interference coordination (ICIC) based on Q learning with Pico BSs (PBSs) and a macro BS (MBS)
as the learning agents. In the case of time-domain ICIC, the action performed by each PBS is to select the bias value for cell range expansion and transmit power on each resource block (RB), and the action of the MBS is only to choose the transmit power. The state of each agent is defined by a tuple of variables, each of which is related to the SINR
condition of each UE, while the received cost of each agent is defined to meet the total transmit power constraint and make the SINR of each served UE approach a target value.
In each iteration of the algorithm, each PBS first selects an action leading to the smallest Q value for the current state, and then the MBS selects its action
in the same way.
While in the case of frequency-domain ICIC, the only difference lies in the action definition of Q learning.
Utilizing Q learning, the Pico and Macro tiers can autonomously optimize system performance with a little coordination.
In \cite{42}, authors use Q learning to optimize the transmit power of SBSs in order to reduce the interference on each RB.
With learning capabilities, each SBS does not need to acquire the strategies of other players
explicitly. In contrast, the experience is preserved in the Q values during the interaction with other SBSs.
To apply Q learning, the state of each SBS is represented as a binary variable that indicates whether the QoS requirement is violated,
and the action is the selection of power levels.
When the QoS requirement is met, the reward is defined as the achieved, instantaneous rate, which equals to zero otherwise.
The simulation result demonstrates that Q-learning can increase the long-term expected data rate of SBSs.

Another important scenario requiring power control is the CR scenario.
In \cite{53}, distributed Q learning is conducted to manage aggregated interference generated by
multiple CRs at the
receivers of primary (licensed) users, and the secondary BSs are taken as learning agents.
The state set defined for each agent is composed of a binary variable indicating whether the secondary system generates excess interference to primary receivers,
the approximate distance between the secondary user (SU) and protection contour, and the transmission power corresponding to the current SU.
The action set of the secondary BS is the set of power levels that can be adopted with the cost
function designed to limit the interference at the primary receivers.
Taking into account that the agent cannot always obtain an accurate observation of the interference indicator,
authors then discuss two cases, namely complete information and partial information, and handle the latter
by involving belief states in Q learning.
Moreover, two different ways of Q value representation are discussed utilizing look-up tables and neural networks,
and the memory, as well as, computation overheads are also examined.
Simulations show that the proposed scheme can allow agents to learn a series of optimization policies that will keep the aggregated
interference under a desired value.

In addition, some research utilizes reinforcement learning to achieve the equilibrium state of wireless networks, where
power control problems are modeled as non-cooperative games among multiple nodes.
In \cite{Power+}, the power control of femto BSs (FBSs) is conducted to mitigate the cross-tier interference to macrocell UEs (MUEs). Specifically,
the power control and carrier selection is modeled as a normal form game among FBSs in mixed strategies, and a reinforcement learning
algorithm based on joint utility and strategy estimation is proposed to help FBSs reach Logit equilibrium.
In each iteration of the algorithm, each FBS first selects an action according to its current strategy, and then receives a reward which
equals its data rate if the QoS of the MUE is met and equals to zero otherwise.
Based on the received reward, each FBS updates its utility estimation and strategy by the process proposed in \cite{ju}.
By numerical simulations, it is demonstrated that the proposed algorithm can achieve a satisfactory performance while
each FBS does not need to know any information of the game but its own reward.
It is also shown that
taking identical utilities for FBSs benefits the whole system performance.

In \cite{sun13}, authors model the channel and power level selection of D2D pairs in a heterogeneous cellular
network as a stochastic non-cooperative game.
The utility of each pair is defined by considering the SINR constraint and the difference between its achieved data rate and the cost of power consumption.
To avoid the considerable amount of information exchange among pairs incurred by
conventional multi-agent Q learning,
an autonomous Q learning algorithm is developed based on the estimation of pairs' beliefs about the
strategies of all the other pairs.
Finally, simulation results indicate that the proposal possesses a relatively fast convergence
rate and can achieve near-optimal performance.

\subsubsection{Supervised Learning Based Approaches}


Considering the high complexity of traditional optimization based resource allocation algorithms,
authors in \cite{nnbeam,nnpower} propose to utilize deep NNs to develop power allocation algorithms that can achieve real-time processing.
Specifically, different from the traditional ways of approximating iterative algorithms where each iteration is approximated by a single layer of the NN,
authors in \cite{nnbeam} adopts a generic dense neural network to approximate the classic
WWMSE algorithm for power control in a scenario with multiple transceiver pairs.
Notably, the number of layers and the number of ReLUs and binary units that are needed for achieving a given approximation error are
rigourously analyzed, from which it can be concluded that the approximation error just has a little impact on the size of the deep NN.
As for NN training, the training data set is generated by running the WMMSE algorithm under varying channel realizations,
and channel realizations together with corresponding power allocation results output by the WMMSE algorithm constitute labeled data.
Via simulations, it is demonstrated that the adopted fully connected NN can achieve similar performance but
with much lower computation time compared with the WMMSE algorithm.

While in \cite{nnpower}, a CNN based power control scheme is developed for the same scenario considered in \cite{nnbeam},
where the full channel gain information is normalized and taken as the input of the CNN, while the output is the power allocation vector.
Similar to \cite{nnbeam}, the CNN is firstly trained to approximate traditional WMMSE to guarantee a basic performance,
and then the loss function is further set as a function of spectral efficiency (SE) or energy efficiency (EE).
Simulation result shows that the proposal can achieve almost the same or even higher SE and EE than WMMSE
at a faster computing speed.
Also using NNs for power allocation, authors in \cite{auto-power} adopt an NN architecture
formed by stacking multiple encoders from pre-trained auto-encoders and a pre-trained softmax layer.
The architecture takes the CSI and the location indicators as the input with each indicator representing whether a user is a cell-edge user,
and the output is the resource allocation result.
The training data set is generated by solving a sum rate maximization problem under different CSI realizations
via the genetic algorithm.

\subsubsection{Transfer Learning Based Approaches}

In heterogenous networks, when femtocells share the same radio resources with macrocells, power control is needed to limit the inter-tier interference to MUEs.
However, facing with dynamic environment, it is difficult for femtocells to meet the QoS constraints of MUEs during the entire operation time.
In \cite{Add-1}, distributed Q-learning is utilized for inter-tier interference management, where the femtocells, as learning agents, aim at
optimizing their own capacity while satisfying the data rate requirement of MUEs.
Due to the frequent changes in RB scheduling, i.e., the RB allocated to each UE is different from time-to-time, and
the backhaul latency,
the power control policy learned by femtocells will be no longer useful and can cause the violation of the data rate constraints of MUEs.
To deal with this problem, authors propose to let the MBS inform femtocells about the future RB scheduling, which facilitates
the power control knowledge transfer between different environments, and hence, femtocells can still avoid interference to the MUE, even if
its RB allocation is changed.
In this study, the power control knowledge is represented by the complete Q-table learned for a given RB.
System level simulations demonstrate this scheme can normally work in a multi-user OFDMA network and is superior to the traditional power control algorithm in terms of
the average capacity of cells.

\subsubsection{Lessons Learned}
It can be learned from \cite{24,42,53,Power+,sun13,Add-1} that distributed Q learning
and learning based on joint utility and strategy estimation can both help to develop self-organizing and autonomous
power control schemes for CRNs and heterogenous networks.
Moreover, according to \cite{53}, Q values can be represented in a tabular form or by a neuron network
that have different memory and computation overheads.
In addition, as indicated by \cite{Add-1}, Q learning can be enhanced to make
agents better adapt to a dynamic environment by involving transfer learning
for a heterogenous network.
Following \cite{Power+}, better system performance can be achieved by
making the utility of agents identical to the system's goal.
At last, according to \cite{nnbeam,nnpower,auto-power}, using NNs to approximate traditional high-complexity power allocation algorithms is
a potential way to realize real-time power allocation.

\begin{table*}[!htp]
\caption{Machine Learning Based Power Control}
\scriptsize
\centering
\begin{tabular}{p{1cm}p{2cm}p{4 cm}p{2 cm}p{6 cm}}
\toprule
\textbf{Literature} & \textbf{Scenario}  & \textbf{Objective} & \textbf{Machine learning technique} & \textbf{Main conclusion}\\
\midrule 

\cite{24}  & A heterogenous network with picocells underlaying macrocells &  Achieve a target SINR for each UE under total transmission power constraints  & Two-level Q learning & The algorithm makes the average throughput improve significantly\\
\midrule

\cite{42}  & A small cell network & Optimize the data rate of each SBS & Distributed Q learning & The long-term expected data rates of SBSs are increased\\
\midrule

\cite{53}  & A cognitive radio network & Keep the interference at the primary receivers below a threshold & Distributed Q learning & The proposals outperform
comparison schemes in terms of outage probability\\
\midrule

\cite{Power+}  & A heterogenous network comprised of FBSs and MBSs & Optimize the throughput of FUEs under the QoS constraints of MUEs & Reinforcement learning with joint utility and strategy estimation & The algorithm can converge to the Logit equilibrium, and the spectral efficiency is higher when FBSs take the system performance as their utility\\
\midrule

\cite{sun13}  & A D2D enabled cellular network & Optimize the reward of each D2D pair defined as the difference between achieved data rate and transmit power cost under QoS constraints& Distributed Q learning & The algorithm is proved to converge to the optimal Q values and improves the average throughput significantly\\
\midrule

\cite{66}  & A cognitive radio network & Optimize transmit power level selection to reduce interference & SVM & The proposed algorithm not only achieves a tradeoff between energy efficiency and satisfaction index, but also satisfies the probabilistic interference constraint\\
\midrule

\cite{67}  & A cellular network & Minimize the total transmit power of devices in the network & SVM & The scheme can balance between the chosen transmit power and the user SINR\\
\midrule

\cite{Add-1}& A heterogenous network with femtocells and macrocells & Optimize the capacity of femtocells under the transmit power constraints and QoS constraints of MUEs & Knowledge transfer based Q learning & The proposed scheme works properly in multi-user OFDMA networks and outperforms conventional power control algorithms \\
\midrule

\cite{nnbeam} & A scenario with multiple transceiver pairs coexisting & Optimize system throughput & Densely connected neural networks & The proposal can achieve almost the same performance compared to WMMSE at a faster computing speed\\
\midrule

\cite{nnpower} & A scenario with multiple transceiver pairs coexisting & Optimize the SE and EE of the system & Convolutional neural networks & The proposal can achieve almost the same or even higher SE and EE than WMMSE at a faster computing speed\\
\midrule
\cite{auto-power} & A downlink cellular network with multiple cells & Optimize system throughput& A multi-layer neural network based on auto-encoders &The proposal can successfully predict the solution of the genetic algorithm in most of the cases\\
\bottomrule
\end{tabular}
\end{table*}

\subsection{Machine Learning Based Spectrum Management}

With the explosive increase of data traffic, spectrum shortages have drawn big concerns in the wireless communication community,
and efficient spectrum management is desired to improve spectrum utilization.
In the following, reinforcement learning and unsupervised learning based spectrum management are introduced.

\subsubsection{Reinforcement Learning Based Approaches}

In \cite{sun39}, spectrum management in millimeter-wave, ultra-dense networks is investigated£¬
and temporal-spatial reuse is considered as a method to improve spectrum utilization.
The spectrum management problem is formulated as a non-cooperative game among devices,
which is proved to be an ordinary potential game guaranteeing the existence of Nash equilibrium (NE).
To help devices achieve NE without global information, a novel, distributed Q learning
algorithm is designed, which facilitates devices to learn environments from the individual
reward.
The action and reward of each device are channel selection and channel capacity, respectively.
Different from traditional Q learning where the Q value is defined over state-action pairs,
the Q value in the proposal is defined over actions, and that is each action corresponds to a Q value.
In each time slot, the Q value of the played action is updated as a weighted sum of the current Q value
and the immediate reward, while the Q values of other actions remain the same.
In addition, based on rigorous analysis, a key conclusion is drawn that
less coupling in learning agents can help speed up the convergence of learning.
Simulations demonstrate that the proposal can converge faster and is more stable than several baselines,
and also leads to a small latency.

Similar to \cite{sun39}, authors in \cite{9} focus on temporal-spatial spectrum reuse but with the use of MAB theory.
In order to overcome the high computation cost brought by the centralized channel allocation policy,
a distributed three-stage policy is proposed, in which the goal of the first two stages is to
help SU find the optimal channel access rank, while the third stage, based on MAB,
is for the optimal channel allocation.
Specifically, with
probability 1-$ \varepsilon $, each SU chooses a channel based on the channel access rank and empirical idle probability estimates,
and uniformly chooses a channel at random otherwise.
Then, the SU senses the selected channel and will receive a reward equal to 1 if neither the primary user nor other SUs transmit over this channel.
By simulations, it is shown that the proposal can achieve significantly smaller regrets
than the baselines in the spectrum temporal-spatial reuse scenario, and
the regret is defined as the difference between the total reward of a genie-aided rule and the expected reward of all SUs.

In \cite{sun40}, authors study a multi-objective, spectrum access problem in a heterogenous network.
Specifically, the concerned problem aims at minimizing the received intra/
inter-tier interference at the femtocells and the inter-tier interference from femtocells to eNBs simultaneously
under QoS constraints.
Considering the lack of global and complete channel information, unknown number
of nodes, and so on, the formulated problem is very challenging.
To handle this issue, a reinforcement learning approach based on joint utility and strategy estimation is proposed,
which contains two sequential levels. The purpose of the first level is to identify available spectrum resource for femtocells,
while the second level is responsible for the optimization of resource selection.
Two different utilities are designed for each level, namely the spectrum modeling and spectrum selection utilities.
In addition, three different learning algorithms, including the gradient follower, the modified RothErev, and the modified Bush and Mosteller
learning algorithms, are available for each femtocell.
To determine the action selection probabilities based on the propensity of each action output by the learning process,
logistic functions are utilized, which are commonly used in machine learning to transform the
full-range variables into the limited range of a probability.
Using the proposed approach, higher cell throughput is achieved owing to the significant reduction in
intra-tier and inter-tier interference.
In \cite{sun}, the joint communication mode selection and subchannel allocation of D2D pairs is solved
by joint utility and strategy estimation based reinforcement learning for a D2D enabled C-RAN.
In the proposal, each action of a D2D pair is a tuple of a communication mode and a subchannel.
Once each pair has selected its action, distributed RRH association and power control are executed,
and then each pair receives the system SE as the instantaneous utility, based on which the utility estimation for each
action is updated. Via simulation, it is demonstrated that a near optimal performance can be achieved by properly
setting the parameter controlling the balance between exploration and exploitation.


To overcome the challenges existing in current solutions to spectrum sharing between operators,
an inter-operator proximal spectrum sharing (IOPSS) scheme is presented in \cite{sun43}, in which a BS
is able to intelligently offload users to its neighboring BSs based on spectral proximity.
To achieve this goal, a Q learning framework is proposed, resulting in a self-organizing, spectrally efficient network.
The state of a BS is the experienced load whose value is discretized, while
an action of a BS is a tuple of spectral sharing parameters related to each neighboring BS,
including the number of RBs requiring each neighboring BS to be reserved, the probability of each user
served by the neighboring BS with the strongest SINR, and the reservation proportion of the requested RBs.
The cost function of each BS is related to both the QoE of its users and the change in the number of RBs it requests.
Through extensive simulations with different loads,
the distributed, dynamic IOPSS based Q learning can help mobile network operators provide
users with a high QoE and reduce operational costs.

In addition to adopting classical reinforcement learning methods, a novel reinforcement learning approach involving recurrent neural networks is
utilized in \cite{3} to handle the management of both licensed and unlicensed frequency bands in LTE-U systems.
Specifically, the problem is formulated as a non-cooperative game with SBSs and an MBS as game players.
Solving the game is a challenging task since each SBS may know only a little information about the network, especially in a
dense deployment scenario.
To achieve a mixed-strategy NE, multi-agent reinforcement learning based on echo state networks is proposed, which
are easy to train and can track the state of a network over time.
Each BS is an ESN agent using two ESNs, namely ESN $\alpha$ and ESN $\beta$, to approximate the immediate and the expected utilities, respectively.
The input of the first ESN comprises the action profile of all the other BSs, while the input of the latter is the user association of the BS.
Compared to traditional RL approaches, the proposal can quickly
learn to allocate resources with not much training data.
During the algorithm execution, each BS needs to broadcast only the action currently taken and its optimal action.
The simulation result shows that the proposed approach improves the sum rate of the 50th percentile of users by up to 167\% compared to Q learning.
A similar idea that combines RNNs with reinforcement learning has been adopted by \cite{vrchen} in
a wireless network supporting virtual reality (VR) services.
Specifically, a complete VR service consists of two communication stages. In the uplink, BSs collect tracking information from users,
while BSs transmit three-dimensional images together with audio to VR users in the downlink.
Thus, it is essential for resource block allocation to jointly consider both the uplink and downlink.
To address this problem, each BS is taken as an ESN agent whose actions are its resource block allocation plans for both uplink and downlink.
The input of the ESN maintained by each BS is a vector containing the indexes of the probability distribution
that all the BSs currently use, while the output is a vector of utility values associated with each action,
based on which the BS selects its action.
Simulation results show that the proposed
algorithm yields significant gains, in terms of VR QoS utility.

\subsubsection{Lessons Learned}

First, it is learned from \cite{sun39} that
less coupling in learning agents can help speed up the convergence of distributed reinforcement learning
for a spatial-temporal spectrum reuse scenario.
Second, following \cite{sun}, near-optimal system SE can be achieved for a D2D enabled C-RAN by
joint utility and strategy estimation based reinforcement learning, if the parameter
balancing exploration and exploitation becomes larger with time going by.
Third, as indicated by \cite{3,vrchen}, when each BS is allowed to manage its own spectrum resource,
RNN can be used by each BS to predict its utility, which can help the system reach a stable resource allocation outcome.

\begin{table*}[!htp]
\caption{Machine Learning Based Spectrum Management}
\centering
\scriptsize
\begin{tabular}{p{1cm}p{2cm}p{4 cm}p{2 cm}p{6 cm}}
\toprule
\textbf{Literature} & \textbf{Scenario}  & \textbf{Objective} & \textbf{Machine learning methods} & \textbf{Main conclusion}\\
\midrule 
\cite{sun39}  & An ultra-dense network with millimeter-wave & Improve spectrum utilization with temporal-spatial spectrum reuse & Distributed Q learning &
Less coupling in learning agents can help speed up the convergence of learning\\
\midrule

\cite{9}  & A cognitive radio network & Improve spectrum utilization with temporal-spatial spectrum reuse & Multi-armed Bandit & The scheme has less regret than other methods when temporal-spatial spectrum reuse is allowed\\
\midrule

\cite{sun40}  & A heterogenous network & Reduce inter-tier and intra-tier interference & Reinforcement learning & Higher cell throughput is achieved owing to the significant reduction in intra-tier and inter-tier interference\\
\midrule

\cite{sun}  & A D2D enabled C-RAN & Optimize system spectral efficiency & Joint utility and strategy estimation based learning & The proposal can achieve near optimal performance in a distributed manner\\
\midrule

\cite{sun43}  & Spectrum sharing among multi-operators & Fully reap the benefits of multi-operator spectrum sharing & Q learning & By the proposal, mobile network operators can serve users with high QoE even when all operators' BSs are equally loaded\\
\midrule

\cite{3}  & A wireless network with LTE-U & Optimize network throughput by user association, spectrum allocation and load balancing & Multi-agent reinforcement learning based on ESNs  & The proposed approach improves the system performance significantly, in terms of the sum-rate of the 50th percentile of users, compared with a Q learning algorithm\\
\midrule

\cite{vrchen}  & A wireless network supporting virtual reality &  Maximize the users' QoS & Multi-agent reinforcement learning based on ESNs  & The proposed
algorithm can achieve significant gains, in terms of VR QoS\\
\bottomrule
\end{tabular}
\end{table*}

\subsection{Machine Learning Based Backhaul Management}

In wireless networks, in addition to the management of radio resources like power and spectrum, the management of backhaul links connecting SBSs and MBSs or connecting BSs and the core network is essential as well to achieve better system performance.
This subsection will introduce literatures related to backhaul management based on reinforcement learning.

\subsubsection{Reinforcement Learning Based Approaches}


In \cite{Conf-36}, Jaber et al. propose a backhaul-aware, cell range extension (CRE) method based on RL to adaptively
set the CRE offset value.
In this method, the observed state for each small cell is defined as a value reflecting the violation of its backhaul capacity,
and the action to be taken is the CRE bias of a cell considering whether the backhaul is available or not. The definition of the cost for each small cell intends to maximize the utilization of total backhaul capacity while keeping the backhaul capacity constraint of each cell satisfied.
Q learning is adopted to minimize this cost through an iterative process, and simulation results show
that the proposal relieves the backhaul congestion in macrocells and improves the QoE of users.
In \cite{75}, authors concentrate on load balancing to improve backhaul resource utilization
by learning system bias values via a distributed Q learning algorithm.
In this algorithm, Xu et al. take the backhaul utilization quantified to several levels as the
environment state based on which each SBS determines an action, that is, the bias value. Then,
with the reward function defined as the weighted difference between the backhaul resource utilization and the outage probability for each SBS,
Q learning is utilized to learn the bias value selection strategy, achieving a balance between system-centric performance and user-centric performance.
Numerical results show that this algorithm is able to optimize the utilization of backhaul resources under the promise of guaranteeing user QoS.


Unlike those in \cite{Conf-36} and \cite{75}, authors in \cite{Backhaul+} and \cite{46} model backhaul management
from a game-theoretic perspective. Problems are solved employing an RL approach based on joint utility and strategy estimation.
Specifically, the backhaul management problem is formulated as a minority game in \cite{Backhaul+}, where
SBSs are the players and have to decide whether to download files for predicted requests while serving the urgent demands.
In order to approximate the mixed NE, an RL-based algorithm, which enables each SBS to update its strategy based on only the received utility, is proposed.
In contrast to previous, similar RL algorithms, this scheme is mathematically proved to converge to a unique equilibrium point for the formulated game.
In \cite{46}, MUEs can communicate with the MBS with the help of SBSs serving as relays.
The backhaul links between SBSs and the MBS are heterogenous including both wired and wireless backhaul.
The competition among MUEs is modeled as a non-cooperative game, where their actions are the selections of
transmission power, the assisting SBSs, and rate splitting parameters.
Using the proposed RL approach, coarse correlated equilibrium of the game is reached.
In addition, it is demonstrated that the proposal achieves better average throughput and delay for the MUEs
than existing benchmarks do.


\subsubsection{Lessons Learned}
It can be learned from \cite{Conf-36} that Q learning
can help with alleviating backhaul congestion and improving the QoE of users
by autonomously adjusting CRE parameters.
Following \cite{75}, when one intends to balance between
system-centric performance and user-centric performance,
the reward fed back to the Q learning agent
can be defined as a weighted difference between them.
As indicated by \cite{Backhaul+},
joint utility and strategy estimation based learning
can help achieve a
balance between downloading files potentially requested in the future
and serving current traffic.
According to \cite{46},
distributed reinforcement learning facilitates UEs
to select heterogenous backhaul links for a heterogenous network scenario
with SBSs acting as relays for MUEs, which results in an improved rate and delay.


\begin{table*}[!t]
\caption{Machine Learning Based Backhaul Management}
\centering
\scriptsize
\begin{tabular}{p{1cm}p{2cm}p{4 cm}p{2 cm}p{6 cm}}
\toprule
\textbf{Literature} & \textbf{Scenario}  & \textbf{Objective} & \textbf{Machine learning methods} & \textbf{Main conclusion}\\
\midrule 
\cite{Conf-37}  & A cellular network & Minimize energy consumption in low traffic scenarios based on backhaul link selection & Q learning & The total energy consumption can be reduced by up to 35\% with marginal QoS compromises \\
\midrule

\cite{Conf-30}  & A cellular network & Optimize joint access and in-band backhaul & Joint utility and strategy estimation based RL & The proposed approach improves system performance by 40\% under completely autonomous operating conditions when compared to benchmark approaches \\
\midrule

\cite{75}  & A cellular network & Maximize the backhaul resource utilization of SBSs and minimize the outage probability of users & Q learning  & The proposed approach effectively utilizes the backhaul resource for load balancing\\
\midrule

\cite{46}  & A cellular network & Improve the throughput and delay of MUEs under heterogeneous backhaul & Joint utility and strategy estimation based RL & The proposed scheme can significantly improve the overall performance \\
\midrule

\cite{Conf-34}  & A cellular network & Maximize system-centric and user-centric performance indicators & Q learning & The proposed scheme shows considerable improvement in users' QoE when compared to state-of-the-art user-cell association schemes \\
\midrule

\cite{Conf-36}  & A cellular network & Optimize cell range extension with backhaul awareness & Q learning & The proposed approach alleviates the backhaul congestion of the macrocell and improves user QOE \\
\midrule


\cite{Backhaul+}  & A cellular network & Minimize the traffic conflict in the backhaul & Joint utility and strategy estimation based RL & The proposed scheme is proved to converge to an approximate version of Nash equilibrium  \\

\bottomrule
\end{tabular}
\end{table*}

\subsection{Machine Learning Based Cache Management}

Due to the proliferation of smart devices and intelligent applications, such as augmented reality, virtual reality, ubiquitous social networking, and IoT,
wireless communication systems have experienced a tremendous data traffic increase over the past couple of years.
Additionally, it has been envisioned that the cellular network will produce about 30.6 exabytes data per month by 2020 \cite{S-8}.
Faced with the explosion of data demands, the caching paradigm is introduced for the future wireless network to shorten latency and alleviate the transmission
burden on backhaul \cite{huan}.
Recently, many excellent research studies have adopted ML techniques to manage cache resource with great success.

\subsubsection{Reinforcement Learning Based Approaches}

Considering the various spatial and temporal content demands among different small cells,
authors in \cite{benecache} develop a decentralized caching update scheme based on joint utility-and-strategy-estimation RL.
With this approach, each SBS can optimize a caching probability distribution over content classes using only the
received instantaneous utility feedback. In addition, by doing weighted sum of the caching strategies of each SBS and the cloud,
a tradeoff between local content popularity and global popularity can be achieved.
In \cite{weicache}, authors also focus on distributed caching design.
Different from \cite{benecache}, BSs are allowed to cooperate with each other in the sense that each BS can get the locally missing content
from other BSs via backhaul, which can be a more cost-efficient solution, and meanwhile D2D offloading is also considered to
improve the cache utilization. Then, to minimize system transmission cost, a distributed Q learning algorithm is utilized.
For each BS, the content placement is taken as the observed state, and the adjustment of cached contents is taken as the action.
The convergence of the proposal is proved by utilizing the sequential
stage game model, and its superior performance is verified via simulations.

Instead of adopting traditional RL approaches,
in \cite{S-1}, authors propose a novel framework based on DRL for a connected vehicular network to orchestrate computing, networking, and cache resource. Particularly, the DRL agent decides which BS is assigned to the vehicle and whether to cache the requested content at the BS.
Simulation results reveal that, by utilizing DRL, the system gains much better performance compared to approaches in existing works.
In addition, authors in \cite{drlcache} investigates a cache update algorithm based on Wolpertinger DRL architecture for a single BS.
Concretely, the request frequencies of each file over different time durations and the current file requests
from users constitute the input state, and the action decides whether to cache the requested
content. The proposed scheme is compared with several traditional cache update schemes including Least Recently Used and Least Frequently Used,
and it is shown that the proposal can raise both short-term cache hit rate and long-term cache hit rate.

\subsubsection{Supervised Learning Based Approaches}

In order to develop an adaptive caching scheme, extreme learning machine (ELM) \cite{elmcache} has been employed
to estimate content popularity.
Hereafter, mixed-integer linear programming is used to compute the content placement.
Moreover, a simultaneous perturbation stochastic approximation method is proposed to reduce
the number of neurons for ELM, while guaranteeing a certain level of prediction accuracy.
Based on real-world data, it is shown that the proposal can improve both the QoE of users and network performance.


In \cite{chenjsac}, the joint optimization of content placement, user association, and unmanned aerial vehicles' positions
is studied, aiming at minimizing the total transmit power of UAVs while satisfying the requirement of user
quality of experience.
To solve the formulated problem that focuses on a whole time duration, it is essential to
predict user content request distribution. To this end, an echo state network (ESN), a kind of recurrent neural networks,
is employed, which can quickly learn the distribution based on not much training data.
Specifically, the input of the ESN is an vector consisting of the user context information like gender and device type,
while the output is the vector of user content request probabilities.
An idea similar to that in \cite{chenjsac} is adopted in \cite{chencran} to optimize the contents cached at RRHs and the BBU pool
for a cloud radio access network.
First, an ESN based algorithm is introduced to enable the BBU pool to learn
each user¡¯s content request distribution and mobility pattern, and then
a sublinear algorithm is proposed to determine which content to cache.
Moreover, authors have derived
the ESN memory capacity for a periodic input.
By simulation, it is indicated that the proposal improves sum effective capacity
by 27.8\% and 30.7\%, respectively, compared to random caching with clustering and random caching
without clustering.

Different from \cite{elmcache,chenjsac,chencran} predicting the content popularity using neural networks directly,
authors in \cite{tonycache} propose a scheme integrating 3D CNN for video generic feature
extraction, SVM for generating representation vectors of videos, and then a regression model for
predicting the video popularity by taking the corresponding representation vector as input.
After the popularity of each video is obtained, the optimal portion of each video cached at the BS is
derived to minimize the backhaul load in each time period. The advantage of the proposal
lies in the ability to predict the popularity of new uploaded videos with no statistical
information required.

\subsubsection{Transfer Learning Based Approaches}

Generally, content popularity profile plays key roles in deriving efficient caching policies, but its estimation with high accuracy
suffers from a long time incurred by collecting user file request samples.
To overcome this issue, authors in \cite{poorcache}
involve the idea of transfer learning by integrating the file request samples from the social network domain
into the file popularity estimation formula.
By theoretical analysis, the
training time is expressed as a function of the ``distance`` between the probability distribution of the requested files and
that of the source domain samples.
In addition, transfer learning based approaches are also adopted
in \cite{tlcache1} and \cite{tlcache2}.
Specifically, although collaborative filtering (CF) can be utilized
to estimate the file popularity matrix, it faces the problem of data sparseness.
Hence, authors in \cite{tlcache1} propose a transfer learning based CF approach to
extract collaborative social behavior information from the
interaction of D2D users within a social community (source domain), which
improves the estimation of the (large-scale) file popularity matrix in the target domain.
In \cite{tlcache2}, a transfer learning based caching scheme is developed, which is executed
at each SBS.
Particularly, by using the contextual information like users' social ties
that are acquired from D2D interactions (source domain),
cache placement at each small cell is carried out,
taking estimated content popularity, traffic load and backhaul capacity into account.
Via simulation, it is demonstrated that the proposal can well deal with data sparsity and cold-start
problems, which results in significant enhancement in users' QoE.

\subsubsection{Lessons Learned}

It can learned from \cite{elmcache,chenjsac,chencran} that
content popularity profile can be accurately
predicted or estimated by
supervised learning like recurrent neural networks and extreme learning machine,
which is useful for the cache management problem formulation.
Second, based on \cite{poorcache,tlcache1,tlcache2},
involving the content request information from other domains, such as the social network domain,
can help reduce the time needed for popularity estimation.
At last, when the file popularity profile is difficult to acquire,
reinforcement learning is an effective way to directly optimize the caching policy,
as indicated by \cite{benecache,weicache,drlcache}.

\begin{table*}[!t]
\caption{Machine Learning Based Cache Management}
\centering
\scriptsize
\begin{tabular}{p{1cm}p{2cm}p{4 cm}p{2 cm}p{6 cm}}
\toprule
\textbf{Literature} & \textbf{Scenario}  & \textbf{Optimization objective} & \textbf{Machine learning methods} & \textbf{Main conclusion}\\
\midrule  
\cite{benecache} & A small cell network & Minimize latency & Joint utility and strategy estimation based RL &  The proposal can achieve 15\% and 40\% gains compared to various baselines\\
\midrule

\cite{weicache} & A D2D enabled cellular network & Minimize system transmission cost & Distributed Q learning &  The proposal outperforms traditional caching strategies including LRU and LFU\\
\midrule

\cite{S-1}  & A vehicular network & Enhance network efficiency and traffic control &  Deep reinforcement learning & The proposed scheme can significantly improve the network performance \\
\midrule

\cite{drlcache} & A scenario with a single BS & Maximize the long-term cache hit rate & Deep reinforcement learning & The proposal can achieve improved short-term cache hit rate and long-term cache hit rate\\
\midrule

\cite{poorcache} & A small cell network & Minimize the latency caused by the unavailability of the requested file & Transfer learning & The training time for popularity distribution estimation can be reduced by transfer learning\\
\midrule

\cite{elmcache} & A cellular network & Improve the users' quality of experience and reduce network traffic & Extreme learning machine & The QoE of users and network performance can be improved compared with industry standard caching schemes\\
\midrule

\cite{chenjsac} & A cloud radio access network with UAVs & Minimize the transmit power of UAVs and meanwhile satisfy the QoE of users & Echo state networks & The proposal achieves significant gains compared to baselines without cache and UAVs\\
\midrule

\cite{chencran} & A cloud radio access network & Maximize the long-term sum effective capacity & Echo state networks &  The proposed approach can considerably improves the sum effective capacity\\
\midrule

\cite{tonycache} & A cellular network & Minimize the average backhaul load & 3D CNN, SVM and regressive model &
Content-aware based proactive caching is cost-effective for dealing with the bottleneck of backhaul\\
\bottomrule
\end{tabular}
\end{table*}

\subsection{Machine Learning Based Computation Resource Management}

In \cite{chenxian}, authors investigate a wireless network that provides MEC services, and
a computation offloading decision problem is formulated for a
representative mobile terminal, where multiple BSs are available for computation offloading.
More specifically, the problem takes environmental dynamics into account including
time-varying channel quality and the task arrival and energy status at the mobile device.
To develop the optimal offloading decision policy, a double DQN based learning approach is proposed, which does not need the complete information about network dynamics
and can handle state spaces with high dimension.
Simulation results show that the proposal can improve
computation offloading performance significantly compared with several baseline policies.

\subsubsection{Lessons Learned}

From \cite{chenxian}, it is learned that
DRL based on double DQN can be used to optimize the computation offloading policy without knowing the information about network dynamics,
such as channel quality dynamics, and meanwhile can handle the issue of state space explosion for a wireless network
providing MEC services.
However, authors in \cite{chenxian} only consider a single user.
Hence, in the future, it is interesting to study the computation offloading for the scenario with multiple users based on DRL,
whose offloading decisions can be coupled due to interference and constrained MEC resources.

\begin{table*}[!t]
\caption{Machine Learning Based Computation Resource Management}
\centering
\scriptsize
\begin{tabular}{p{1cm}p{2cm}p{4 cm}p{2 cm}p{6 cm}}
\toprule
\textbf{Literature} & \textbf{Scenario}  & \textbf{Optimization objective} & \textbf{Machine learning methods} & \textbf{Main conclusion}\\
\midrule 
\cite{chenxian} & An MEC scenario with a representative user and multiple BSs & Optimize a long-term utility that is related to task execution delay, task queuing delay, and so on& DRL &  The proposal can improve
computation offloading performance significantly compared with several baseline policies\\
\bottomrule
\end{tabular}
\end{table*}

\subsection{Machine Learning Based Beamforming}

Considering the ever-increasing QoS requirements and the need for real-time processing in practical systems,
authors in \cite{knnbeam} propose a supervised learning based resource allocation framework to
quickly output the optimal or a near optimal resource allocation solution for the current scenario.
Specifically, the data related to historical scenarios is collected and the feature vector is extracted for
each scenario. Then, the optimal or near optimal resource allocation plan can be searched off-line by taking
the advantage of cloud computing. After that, those feature vectors with the same resource allocation solution
are labeled with the same class index. Up to now, the remaining task to determine resource allocation
for a new scenario is to identify the class of its corresponding feature vector, and that is the resource allocation
problem is transformed into a multi-class classification problem, which can be handled by supervised learning.
To make the application of the proposal more intuitive, an example using KNN to optimize beam allocation in a
single cell with multiple users is shown, and simulation results show an improvement in terms of
sum rate compared to a state-of-the-art method.

\subsubsection{Lessons Learned}
As indicated by \cite{knnbeam},
resource management in wireless networks can be transformed into a supervised, classification task,
where the labeled data set is composed of feature vectors representing different scenarios and their corresponding classes.
The feature vectors belonging to the same class correspond to the same resource allocation solution.
Then, various machine learning techniques for classification can be applied to determine the resource allocation for
a new scenario.
When the classification algorithm is with low-complexity,
it is possible to achieve near real-time resource allocation.
At last, it should be highlighted that the
key to the success of this framework lies in the proper feature vector construction for the communication scenario and the design of
low-complexity multi-class classifiers.

\begin{table*}[!t]
\caption{Machine Learning Based Beamforming Design}
\centering
\scriptsize
\begin{tabular}{p{1cm}p{2cm}p{4 cm}p{2 cm}p{6 cm}}
\toprule
\textbf{Literature} & \textbf{Scenario}  & \textbf{Optimization objective} & \textbf{Machine learning methods} & \textbf{Main conclusion}\\
\midrule 
\cite{knnbeam} & A scenario with a single cell and multiple users & Optimize system sum rate & KNN &  The proposal can further raise system performance compared to a state-of-the-art approach\\
\bottomrule
\end{tabular}
\end{table*}

\section{Machine Learning Based Networking}  

With the rapid growth of data traffic and the expansion of the network, networking in future wireless communications requires more efficient solutions.
In particular, the imbalance of traffic loads among heterogenous BSs needs to be addressed, and meanwhile,
wireless channel dynamics and newly emerging vehicle networks both incur a big challenge for traditional networking algorithms that are mainly designed for static networks.
To overcome these issues, research on ML based user association, BS switching control, routing, and clustering
has been conducted.

\subsection{Machine Learning Based BS Association} 

\subsubsection{Reinforcement Learning Based Approaches}

In the vehicle network, the introduction of economical SBSs greatly reduces the network operation cost.
However, proper association schemes between vehicles and BSs are needed for load balancing among SBSs and MBSs. Most previous algorithms
often assume static channel quality, which is not feasible in the real world.
Fortunately, the traffic flow in vehicular networks possesses spatial-temporal regularity. Based on this observation,
Li et al. in \cite{U-1} propose an online reinforcement learning approach (ORLA) for a vehicular network.
The proposal is divided into two learning phases: initial reinforcement learning and history-based reinforcement learning.
In the initial learning model,
the vehicle-BS association problem is seen as a multi-armed bandit problem, where the action of each BS is the decision on the association with
vehicles and the reward is defined to minimize the deviation of the data rates of the vehicles from the average rate of all the vehicles.
In the second learning phase, considering the spatial-temporal regularities of vehicle networks,
the association patterns obtained in the initial RL stage
enable the load balancing of BSs through history-based RL when the environment dynamically changes.
Specifically, each BS will calculate the similarity between the current environment and each historical
pattern, and the association matrix is output based on the historical association pattern.
Compared with the max-SINR scheme and distributed dual decomposition optimization, the proposed ORLA reaches the minimum load variance of multiple cells.



Besides the information related to SINR, backhaul capacity constraints and diverse attributes related to the QoE of users
should also be taken into account for user association.
In \cite{U-Conf-1}, authors propose a distributed, user-centric, backhaul-aware user association scheme based on fuzzy Q learning
to enable each cell to autonomously maximize its throughput under backhaul capacity constraints and user QoE constraints.
More concretely, each cell broadcasts a set of bias values to guide users to associate with preferred cells,
and each bias value reflects the capability to satisfy a kind of performance metrics like throughput and resilience.
Using fuzzy Q learning, each cell tries to learn the optimal bias values for each of the fuzzy rules through
iterative interaction with the environment.
The proposal is shown to outperform the traditional Q learning solutions in both computational efficiency and system capacity.

In \cite{ha}, an uplink user association
problem for energy harvesting devices in an ultra-dense small cell network is studied.
Faced with the uncertainty of channel gains, interference, and/or user
traffic, which directly affects the probability to receive a positive reward,
the association problem is formulated as an MAB problem, where each device selects an SBS for transmission in each
transmission round.

Considering the trend to integrate cellular-connected UAVs in future wireless networks,
authors in \cite{waliduav} study a joint optimization of optimal paths, transmission power levels, and cell associations
for cellular-connected UAVs to minimize the wireless latency of UAVs and their interference
to the ground network. The problem is modeled as a dynamic game with UAVs as game players,
and an ESN based deep reinforcement learning approach is proposed to solve the game.
Specifically, the deep ESN after training enables each UAV to decide an action based on
the observation of the network state.
Once the approach converges, a subgame perfect Nash equilibrium is reached.

\subsubsection{Collaborative Filtering Based Approaches}

Historical network information is beneficial for obtaining the service capabilities of BSs, from which the similarities between the preferences of users selecting the cooperating BSs will also be derived. Meng et al. in \cite{U-3} propose an association scheme considering both the historical QoS information of BSs and user social interactions in heterogeneous networks. This scheme contains a recommendation system composed of the rating matrix, UEs, BSs, and operators, which is based on CF. The rating matrix formed by the measurement information received by UEs from their connecting BSs is the core of the system. The voice over internet protocol service is used as an example to describe the network recommendation system in this scheme. The E-model proposed by ITU-T is used to map the SNR, delay, and packet loss rate measured by UEs in real time to an objective mean opinion score of QoS as a rating, thus generating the rating matrix. With the ratings from UEs, the recommendation system guides the user association through the user-oriented neighborhood-based CFs.
Simulation shows that this scheme needs to set a moderate expectation value for the recommendation system to converge within a minimum number of iterations. In addition, this scheme outperforms selecting the BS with the strongest RSSI or QoS.

\subsubsection{Lessons Learned}

First, based on \cite{U-1}, utilizing the spatial-temporal regularity of traffic flow in vehicular
networks helps develop an online association scheme that contributes to a small load variance of multiple cells.
Second, it can be learned from \cite{U-Conf-1} that fuzzy Q learning
can outperform Q learning when the state space of the communication system is continuous.
Third, according to \cite{ha}, the tradeoff between the exploitation of learned association knowledge and the exploration of unknown situations should be carefully
made for MAB based association.

\begin{table*}[!t]
\caption{Machine Learning Based User Association}
\scriptsize
\centering
\begin{tabular}{p{1cm}p{2cm}p{4 cm}p{2 cm}p{4 cm}}
\toprule
\textbf{Literature} & \textbf{Scenario}  & \textbf{Optimization objective} & \textbf{Machine learning methods} & \textbf{Main conclusion}\\
\midrule 

\cite{U-1}  & A vehicular network & Optimize the user association strategy through learning the temporal dimension regularities in the network & Reinforcement learning & The proposed scheme can well balance the traffic load \\
\midrule

\cite{U-3}  & A heterogeneous network & Optimize association between UEs and BSs considering multiple factors affecting QoS & Collaborative filtering & The proposed scheme achieves satisfaction equilibrium between the profits and costs \\
\midrule

\cite{U-Conf-1}  & A heterogeneous network & Optimize user-cell association in a user-centric and backhaul-aware manner & Fuzzy Q learning & The proposed scheme
improves the users' performance by 12\% at the cost of 33.3\% additional storage memory \\
\midrule

\cite{U-Conf-2}  & A heterogeneous network & Optimize user-cell association through learning the cell range expansion & Q learning & The proposed scheme minimizes the number of UE outages and improves the system throughput  \\
\midrule

\cite{ha} & A small cell network with energy harvesting devices & Guarantee the minimum data rate of each device & MAB &  The proposal is applicable in hyper-dense networks\\
\midrule

\cite{waliduav} & A cellular network with UAV-UEs & Minimize the wireless latency of UAVs and their interference to the ground network  & ESN based DRL &  The altitude of the UAVs greatly affects the optimization of the obejective\\
\bottomrule
\end{tabular}
\end{table*}

\subsection{Machine Learning Based BS Switching} 

Deploying a number of BSs is seen as an effective way to meet the explosive growth of traffic demand. However, much energy can be consumed to maintain the operations of BSs.
To lower energy consumption, BS switching is considered a promising solution by switching off the unnecessary BSs \cite{bs13}.
Nevertheless, there exist some drawbacks in traditional switching strategies.
For example, some do not consider the cost led by the on-off state transition,
while some others assume the traffic loads are constant, which is very impractical.
In addition, those methods highly rely on precise, prior knowledge about the environment which is hard to collect.
Facing these challenges, some researchers have revisited BS switching problems from the perspective of ML.

\subsubsection{Reinforcement Learning Based Approaches}

In \cite{bs2}, an actor-critic learning based method is proposed, which avoids the need for prior environmental knowledge.
In this scheme, BS switching on-off operations are defined as the actions of the controller with the traffic load as the state, aiming at minimizing overall energy consumption.
At a given traffic load state, the controller chooses a BS switching action in a stochastic way based on policy values.
After executing a switching operation, the system will transform into a new state and the energy cost of the former state is calculated.
When the energy cost of the executed action is smaller than those of other actions, the controller will update the policy value to enable this action to be more likely to be selected, and vice versa. By gradually communicating with the environment, an optimal switching strategy is obtained when policy values converge.
Simulation results show that the energy consumption of the proposal is slightly higher than that of the state of the art scheme in which the prior knowledge of the
environment is assumed to be known but difficult to acquire in practice.
Similarly, authors in \cite{bs6} propose a Q learning method to reduce the overall energy consumption,
which defines BS switching operations as actions and the state
as a tuple of the user number and the number of active SBSs.
After a switching action is chosen at a given state, the reward,
which takes into account both energy consumption and transmission gains,
can be obtained. After that, with the calculated reward, the system updates the corresponding Q value.
This iterative process goes on until all the Q values converge.

Although the above works have achieved good performance, the power cost incurred by the on-off state transition of BSs is not taken into account.
To make results more rigorous, authors in \cite{bs1} include the transition power in the cost function,
and propose a Q learning method with the action defined as a pair of thresholds named as the upper user threshold and the lower user threshold.
When the number of users in a small cell is higher than the upper threshold, the small cell will be switched on,
while the cell will be switched off once the number of users is less than the lower threshold.
Simulation results show that this method can avoid frequent BS on-off state transitions, thus saving energy consumption.
In \cite{tangjian}, a more advanced power consumption optimization framework based on DRL is
proposed for a downlink cloud radio access network, where the power consumption caused by the on-off state transition of RRHs is also considered.
Simulation results reveal that the proposed framework can achieve much power saving with the satisfaction of user
demands and adaptation to dynamic environments.


Moreover, fuzzy Q learning is utilized in \cite{bs10} to find the optimal sensing probability of the SBS, which directly impacts its on-off operations.
Specifically, an SBS operates in sleep mode when there is no active users to serve, and then it wakes up randomly to sense MUE activity.
Once the activity of an MUE is detected, the SBS goes back to active mode.
By simulation, it is demonstrated that the proposal can well handle user density
fluctuations and can improve energy efficiency while guaranteeing network capacity and coverage probability.


\subsubsection{Transfer Learning Based Approaches}

Although the discussed studies based on RL show favorable results, the performance and convergence time can be further improved by
fully utilizing prior knowledge.
In \cite{bs3}, which is an enhanced study of \cite{bs2},
authors integrate transfer learning (TL) into actor-critic reinforcement learning and propose a novel BS switching control algorithm.
The core idea of the proposal is that the controller can utilize
the knowledge learned in historical periods or neighboring regions to help find the optimal BS switching operations.
More specifically, the knowledge refers to the policy values in actor-critic reinforcement learning.
The simulation result shows that combining RL with transfer learning outperforms the method only using RL, in terms of both energy saving and convergence speed.

Though some TL based methods have been employed to develop BS sleeping strategies, WiFi network scenarios have not been covered.
Under the context of WiFi networks, the knowledge of the real time data, gathered from the APs related to the present environment,
is utilized for developing switching on-off policy
in \cite{bs5}, where the actor-critic algorithm is used.
These works have indicated that
TL can offer much help in finding optimal BS switching strategies,
but it should be noted that TL may lead to a negative influence in the network, since
there are still differences between the source task and the target task.
To resolve this problem, authors in \cite{bs3} propose to diminish the impact of the prior knowledge on decision making with time going by.

\subsubsection{Unsupervised Learning Based Approaches}

K-means, which is a kind of unsupervised learning algorithm, can help enhance BS switching on-off strategies.
In \cite{bs12}, based on the similarity of the location and traffic load of BSs, K-means clustering is used to group BSs into
different clusters, within each of which the interference is mitigated by allocating orthogonal resources among communication links, and
the traffic of off-BSs can be offloaded to on-BSs.
Simulation shows that involving K-means results in a lower average cost per BS when the number of users is large.
In \cite{bs15}, by applying K-means, different values of RSRQ are grouped into different clusters. Consequently, the users will be grouped into clusters, which depends
on their corresponding RSRQ values. After that, the cluster information is considered as a part of the system state in Q learning to find the optimal BS switching strategy.
With the help of K-means, the proposed method achieves lower average energy consumption than the method without K-means.

\subsubsection{Lessons Learned}

First, according to \cite{bs2},
the actor-critic learning based method enables BSs to make wise switching decisions without the need for
knowledge about the traffic loads within the BSs.
Second, following \cite{bs1,tangjian}, the power cost incurred by BS switching should be involved
in the cost fed back to the reinforcement learning agent, which makes the energy consumption
optimization more reasonable.
Third, as indicated by \cite{bs3}, integrating transfer learning into actor-critic learning based BS switching can achieve better performance at the beginning as well as faster convergence compared to the traditional actor-critic learning.
At last, based on \cite{bs12,bs15}, by properly clustering BSs and users by K-means before optimizing BS on-off states, better performance can be gained.

\begin{table*}[!t]
\caption{Machine Learning Based BS Switching}
\centering
\scriptsize
\begin{tabular}{p{1cm}p{2cm}p{4 cm}p{2 cm}p{6 cm}}
\toprule
\textbf{Literature} & \textbf{Scenario}  & \textbf{Optimization objective} & \textbf{Machine learning methods} & \textbf{Main conclusion}\\
\midrule 
\cite{bs1}  & A Hetnet & Minimize the network energy consumption & Q learning & The proposal can avoid the frequent on-off state transitions of SBSs \\
\midrule
\cite{bs2}  & A cellular network & Improve energy efficiency & Actor-critic reinforcement learning & The proposed method achieves similar energy consumption under dynamic traffic loads compared with the method which has a full traffic load knowledge \\
\midrule
\cite{bs3}  & A cellular network & Improve energy efficiency & Transfer learning based actor-critic  & Transfer learning based RL outperforms classic RL method in energy saving\\
\midrule
\cite{bs4}  & A Hetnet & Optimize the trade-off between total delay experienced by users and energy savings & Transfer learning based actor-critic  & Transfer learning based RL outperforms classic RL method in energy saving\\
\midrule
\cite{bs5}  & A WiFi network & Improve energy efficiency & Transfer learning based actor-critic  & Transfer learning based RL can achieve higher energy efficiency \\
\midrule
\cite{bs6}  & A Hetnet & Improve energy efficiency & Q learning & Q learning based approach outperforms the traditional static policy\\
\midrule
\cite{bs8}  & A Hetnet & Improve drop rate, throughput, and energy efficiency & Q learning & The proposed multi-agent Q learning method outperforms the traditional greedy method \\
\midrule
\cite{bs9}  & A green wireless network & Maximize energy saving & Q learning & Different learning rates and discount factors of Q learning will significantly influence the energy consumption\\
\midrule
\cite{tangjian}& A cloud radio access network & Minimize system power consumption & Deep reinforcement learning & The proposed framework can achieve much power saving with the satisfaction of user demands and can adapt to dynamic environment\\
\midrule
\cite{bs10}  & A Hetnet & Improve energy efficiency & Q learning &The proposed method improves energy efficiency while maintaining network capacity and coverage probability\\
\midrule
\cite{bs12}  & A small cell network & Reduce energy consumption & K-means & Proposed clustering method reduces the overall energy consumption \\
\midrule
\cite{bs15}  & An opportunistic mobile broadband network & Optimize the spectrum allocation, load balancing, and energy saving  & Q learning, k-means, and transfer learning & The proposed method achieves lower average energy consumption than the method without K-means \\
\bottomrule
\end{tabular}
\end{table*}

\subsection{Machine Learning Based Routing} 


To fulfill stringent traffic demands in the future, many new RAN technologies continuously come into being, including C-RANs, CR networks (CRNs) and ultra-dense networks. To realize effective networking in these scenarios, routing strategies play key roles.
Specifically, by deriving proper paths for data transmission, transmission delay and other types of performance can be optimized.
Recently, machine learning has emerged as a breakthrough for providing efficient routing protocols to enhance the overall network performance \cite{S-23}.
In this vein, we provide a vivid summarization on novel machine learning based routing schemes.

\subsubsection{Reinforcement Learning Based Approaches}

To overcome the challenge incurred by dynamic channel availability in CRNs,
authors in \cite{S-21} propose a clustering and reinforcement learning based multi-hop routing scheme
, which provides high stability and scalability.
Using Q learning, the availability of the bottleneck channel along the route can be well estimated, which guides the routing node selection.
Specifically, the source node maintains a Q table, where each Q value corresponds to a pair composed of a destination node and the next-hop node.
After Q values are learned and given the destination, the source node chooses the next-hop node with the largest Q value.

Also focusing on multi-hop routing in CRNs, two different routing schemes based on reinforcement learning,
namely traditional RL scheme and RL-based scheme with average Q value,
are investigated in \cite{S-22}.
In both schemes, the definitions of the action and state are the same as those in \cite{S-21},
and the reward is defined as the channel available time of the bottleneck link.
Compared to traditional RL scheme, RL-based scheme with average Q value
can help with selecting more stable routes.
The superior performance of these two schemes is verified by implementing a test bed
and comparing with a highest-channel route selection scheme.

In \cite{S-31}, authors study the influences of several network characteristics, such as network size, on the performance of
Q learning based routing for a cognitive radio ad hoc network.
It is found that network characteristics have slight impacts on the end-to-end delay and packet loss rate
of SUs.
In addition, reinforcement learning is also a promising paradigm for developing routing protocols for the unmanned robotic network.
Specifically, to save network overhead in high-mobility scenarios, a Q learning based geographic routing strategy is introduced in \cite{S-27},
where each state represents a mobile node and each action defines a routing decision.
A characteristic of the Q learning adopted in this paper is the novel design of the reward that
incorporates packet travel speed.
Simulation using NS-3 confirms a better packet delivery ratio but with a lower network overhead compared to existing methods.

\subsubsection{Supervised Learning Based Approaches}

In \cite{S-20}, a routing scheme based on DNNs is developed, which enables each router in heterogenous networks
to predict the whole path to the destination. More concretely, each router trains a DNN to predict the next proper router for
each potential destination using the training data generated by following Open Shortest Path First (OSPF) protocol,
and the input and output of the DNN is the traffic patterns of all the routers and the index of the next router, respectively.
Moreover, instead of training all the weights of the DNN at the same time, a greedy layer-wise
training method is adopted.
By simulations, lower signaling overhead and higher throughput is observed compared with OSPF routing strategy.

To improve the routing performance for the wireless backbone,
deep CNNs are exploited in \cite{S-29}, which can learn from the experienced congestion.
Specifically, a CNN is constructed for each routing strategy, and the CNN takes traffic pattern information collected from routers, such as traffic generation rate,
as input to predict whether the corresponding routing strategy can cause congestion.
If yes, the next routing strategy will be evaluated until it is predicted that there will be no congestion.
Meanwhile, it should be noted that these constructed CNNs are trained in an on-line manner with the training data set continuously updated,
and hence the routing decision becomes more accurate.

\subsubsection{Lessons Learned}

First, according to \cite{S-21,S-22}, when one applies Q learning to route selection in CRNs,
the reward feedback can be set as a metric representing the quality of the bottleneck
link along the route, such as the channel available time of the bottleneck link.
Second, following \cite{S-31}, network characteristics, such as network size,
have limited impacts on the performance of Q learning based routing for a cognitive radio ad hoc network,
in terms of end-to-end delay and packet loss rate of secondary users.
Third, from \cite{S-20},
it can be learned that DNNs can be trained using the data generated by OSPF routing protocol,
and the resulting DNN model based routing can achieve lower signaling overhead and higher throughput.
At last, as discussed in \cite{S-29}, OSPF can be substituted by CNN based routing,
which is trained in an online fashion and can avoid past, fault routing decisions compared
to OSPF.

\begin{table*}[!t]
\caption{Machine Learning Based Routing Strategy}
\centering
\scriptsize
\begin{tabular}{p{1cm}p{2cm}p{4 cm}p{2 cm}p{6 cm}}
\toprule
\textbf{Literature} & \textbf{Scenario}  & \textbf{Objective} & \textbf{Machine learning methods} & \textbf{Main conclusion}\\
\midrule 
\cite{S-20}  & A HetNet & Improve network performance by designing a novel routing strategy & Dense neural networks & Lower signaling overhead and higher throughput is observed compared to OSPF routing strategy\\
\midrule
		
\cite{S-21}  & A CRN & Overcome the challenges faced by multi-hop routing in CRNs & Q learning & The SUs' interference to PUs is minimized, and more stable routes are selected\\
\midrule
		
\cite{S-22}  & A CRN & Choose routes with high QoS & Q learning & RL based approaches can achieve a higher throughput and packet delivery ratio compared to highest-channel route selection approach\\
\midrule
										
\cite{S-27}  & An unmanned robotic network & Mitigate the network overhead requirement of route selection &  Q learning & A better packet delivery ratio is achieved by the proposal but with a lower network overhead\\
\midrule
				
\cite{S-29}  &  Wireless backbone &  Realize real-time intelligent traffic control & Deep convolution neural networks & The proposal can significantly improve the average delay and packet loss rate compared to existing approaches \\
\midrule
		
\cite{S-31}  & A cognitive radio ad hoc network & Minimize interference and operating cost & Q learning & The proposed method can improve the routing efficiency and help minimize interference and operating cost\\

\bottomrule
\end{tabular}
\end{table*}

\subsection{Machine Learning Based Clustering}

In wireless networking, it is common to divide nodes or users into different clusters to conduct some cooperation or coordination within each cluster,
which can further improve network performance. Based on the introduction on ML, it can be seen that the clustering problem can be
naturally dealt with the K-means algorithm, as some papers do.
Moreover, supervised learning and reinforcement learning can be utilized as well.

\subsubsection{Supervised Learning Based Approaches}

To reduce content delivery latency in a cache enabled small cell network, a user clustering based TDMA transmission scheme is proposed in \cite{S-3} under
pre-determined user association and content placement, where the user cluster formation and the time duration to serve each cluster need to be optimized.
Since the number of potential clusters grows exponentially with respect to the number of users served by an SBS,
a DNN is constructed to predict whether each user is in a cluster, which takes
the user channel gains and user demands as input. In this manner, users joining clusters can be quickly identified,
reducing the searching space to get the optimal user cluster formation.

\subsubsection{Unsupervised Learning Based Approaches}

In \cite{Conf-29}, K-means clustering is considered for clustering hotspots in densely populated ares with the goal of maximizing spectrum utilization. In this scheme, the mobile device accessing cellular networks can act as a hotspot to provide broadband access to nearby users called slaves.
The fundamental problem to be solved is to identify which devices play the role of hotspots and the set of users associated with each hotspot.
To this end, authors first adopt a modified version of the constrained K-means clustering algorithm to group
the set of users into different clusters based on their locations, and both the maximum number and minimum number of users in a cluster are set.
Then, the user with the minimum average distance to both the center of the cluster and the BS is selected as the hotspot in each cluster.
After that, a graph-coloring approach is utilized to assign spectrum resource to each cluster, and power and spectrum resource allocation to all
slaves and hotspots are performed later on.
The simulation result shows that the proposal can significantly increase the total number of users that can be served in the system with lower cost and complexity.

In \cite{bs12}, authors use clustering of SBSs to realize the coordination between them.
The similarity between two SBSs considers both their distance and the heterogeneity between their traffic loads,
so that two BSs with shorter distance and higher load difference have more chances to cooperate.
Since the similarity matrix possesses the properties of Gaussian similarity matrix, the SBS clustering problem can be
handled by K-means clustering with each SBS corresponding to an attribute vector composed of its coordinates and traffic load.
By intra-cluster coordination, the number of switched-OFF BSs can be increased by offloading
UEs from SBSs that are switched OFF to active SBSs, compared to the case without clustering and coordination.

\subsubsection{Reinforcement Learning Based Approaches}

In \cite{yufei}, to mitigate the interference in a downlink wireless network containing multiple transceiver pairs operating in the same frequency band,
a cache-enabled opportunistic interference alignment (IA) scheme is adopted.
Facing with dynamic channel state information and content availability at each transmitter, a deep reinforcement learning based approach is developed
to determine communication link scheduling at each time slot, and those scheduled transceiver pairs then perform IA.
To efficiently handle the raw collected data like channel state information, the deep Q network in DRL is built using a convolutional neural network.
Simulation results demonstrate the improved performance of the proposal, in terms of system sum rate and energy efficiency, compared to an existing scheme.

\subsubsection{Lessons Learned}

First, based on \cite{S-3},
DNNs can help identify those users or BSs that are not necessary to join clusters, which
facilitates the searching of optimal cluster formation due to the reduction of the searching space.
Second, following \cite{Conf-29,bs12}, clustering problem can be naturally solved using K-means clustering.
Third, as indicated by \cite{yufei},
deep reinforcement learning can be used to directly select the members forming a cluster in a
dynamic network environment with time-varying CSI and cache states.

\begin{table*}[!t]
\caption{Machine Learning Based Clustering}
\scriptsize
\centering
\begin{tabular}{p{1cm}p{2cm}p{4 cm}p{2 cm}p{6 cm}}
\toprule
\textbf{Literature} & \textbf{Scenario}  & \textbf{Objective} & \textbf{Machine learning technique} & \textbf{Main conclusion}\\
\midrule 
\cite{S-3} & A cache-enabled small cell network & Optimize energy consumption for content delivery & Deep neural networks & By using the designed DNN, the solution quality
can reach around 90\% of the optimum\\
\midrule
\cite{Conf-29}  & A cellular network with devices serving as APs &  Study the performance gain obtained by making some devices provide broadband access to other devices in a densely populated area  & Constrained K-means clustering & Under fixed network resources, the proposed algorithm can
significantly improve the overall performance of network users\\
\midrule
\cite{bs12}  & A small cell network & Minimize the network cost  & K-means clustering &Significant gains can be achieved, in terms of energy expenditure and load reduction, compared to conventional transmission techniques \\
\midrule
\cite{yufei} & A cache-enabled opportunistic IA network & Maximize system sum rate & Deep reinforcement learning & The proposal can achieve improved sum rate and energy efficiency compared to an existing scheme\\
\bottomrule
\end{tabular}
\end{table*}

\section{Machine Learning Based Mobility Management}

In wireless networks, mobility management is a key component to guarantee successful service delivery.
Recently, machine learning has shown its significant advantages in user mobility prediction, handover parameter optimization, and so on.
In this section, machine learning based mobility management schemes are comprehensively surveyed.

\subsection{Reinforcement Learning Based Approaches}


In \cite{M-Conf-2}, authors focus on a two-tier network composed of macro cells and small cells,
and propose a dynamic fuzzy Q learning algorithm for mobility management.
To apply Q learning, the call drop rate together with the signaling load caused by handover
constitutes the system state, while the action space is defined as the set of possible values for the adjustment of handover margin.
The aim is to achieve a tradeoff between the signaling cost incurred by
handover and the user experience affected by call dropping ratio (CDR).
Simulation results show that the proposed scheme is effective in minimizing the number of handovers while keeping the CDR at a desired level.
In addition, Klein et al. in \cite{M-Conf-3} also apply the framework based on fuzzy Q learning to jointly optimize TTT and Hys.
Specifically, the framework includes three key components, namely a fuzzy inference system (FIS), heuristic exploration/exploitation policy (EEP)
and Q learning. The FIS input consists of the magnitude of hysteresis margin and the errors of several KPIs like CDR,
and $\varepsilon$-greedy EEP is adopted for each rule in the rule set.
To show the superior performance of the proposal,
a trend-based handover optimization scheme and a TTT assignment scheme based on velocity estimation
are taken as baselines, and it is observed that the fuzzy Q learning based approach
greatly reduces handover failures compared to the two schemes.

Moreover, achieving load balancing during the handover process is an essential part.
In \cite{M-4}, a
fuzzy-rule based RL system is proposed for small cell networks, which aims to balance traffic load by selecting
transmit power (TXP) and Hys of BSs. Considering that the CBR and the OR
will change significantly when the load in a cell is heavy, the two parameters jointly comprise the observation state.
The adjustments of Hys and TXP are system actions, and the reward is defined such that user satisfaction is optimized.
As a result, the optimal adjustment strategy for Hys and TXP is generated by the Q learning system based on fuzzy rules,
which can minimize the localized congestion of small cell networks.

For the LTE network with multiple SON functions, it is inevitable that optimization conflict exists.
In \cite{M-Conf-4}, authors propose a comprehensive solution for SON functions including handover optimization and load balancing.
In this scheme, the fuzzy Q Learning controller is utilized to adjust the Hys and TTT parameters simultaneously, while the heuristic Diff-Load
algorithm optimizes the handover offset according to load measurements in the cell.
To apply fuzzy Q learning, radio link failure, handover failure and handover ping-pong, which are key performance indicators (KPIs) in the
handover process,
are defined as the input to the fuzzy system.
By LTE-Sim simulation, results show that the proposal enables the joint optimization of the KPIs above.


In addition to Q learning, authors in \cite{1} and \cite{M-Arxiv-1} utilize deep reinforcement learning for mobility management.
In \cite{1}, to overcome the challenge of intelligent wireless network management when a large number of RANs and devices are deployed,
Cao et al. propose an artificial intelligence framework based on DRL.
The framework is divided into four parts: real environment, environment capsule, feature extractor, and policy network.
Wireless facilities in a real environment upload information such as the RSSI to the environment capsule.
Then, the capsule transmits the stacked data to the wireless signal feature extraction part consisting of a CNN and an RNN.
After that, these extracted feature vectors will be input to the policy network that is based on a deep Q network
to select the best action for real network management.
Finally, this novel framework is applied to a seamless handover scenario
with one user and multiple APs. Using the measurement of RSSI as input, the user is guided to select the best AP,
which maximizes network throughput.

In \cite{M-Arxiv-1}, authors propose a two-layer framework to optimize the handover process and reach a balance between the
handover rate and system throughput.
The first step is to apply a centralized control method to classify the UEs according to their mobility patterns with unsupervised learning.
Then, the multi-user handover process in each cluster is optimized in a distributed manner using DRL.
Specifically, the RSRQ received by the user from the candidate BS and the current serving BS index make up the state vector,
and the weighted sum between the average handover rate and throughput is defined as the system reward.
In addition, considering that new state exploration in DRL may start from some unexpected initial points, the performance of UEs will greatly fluctuate.
In this framework, Wang et al. apply the output of the traditional 3GPP handover scheme as training data to initialize the deep Q network
through supervised learning, which can compensate the negative effects caused by exploration at the early stage of learning.

\subsection{Supervised Learning Based Approaches}

Except for the current location of the mobile equipment,
learning an individual's next location enables novel mobile applications and a seamless handover process.
In general, location prediction utilizes the user's historical trajectory information to infer the next position of a user.
In order to overcome the lack of historical information issue, Yu et al. propose a supervised learning based prediction method
based on user activity patterns \cite{M-2}.
The core idea is to first predict the user's next activity and then predict its next location.
Simulation results demonstrate the robust performance of the proposal.

\subsection{Unsupervised Learning Based Approaches}

Considering the impact of RF conditions at cell edge on the setting of handover parameters,
authors in \cite{M-7} propose an unsupervised-shapelets based method to help BSs
be automatically aware of the RF conditions at their cell edge by finding useful patterns from RSRP information reported by users.
In addition, RSRP information can be employed to derive the position at the point of a handover trigger.
In \cite{M-8}, authors propose a modified self-organizing map (SOM) based method to determine whether indoor
users should be switched to another external BS based on their location information.
SOM is a type of unsupervised NN that allows generating a low dimensional output space from the high dimensional discrete input.
The input data in this scheme is RSRP together with the angle of the arrival of the mobile terminal, based on which the real physical location of a user
can be determined by the SOM algorithm.
After that, the handover decision can be made for the user according to pre-defined prohibited and permitted areas.
Through evaluation using the network simulator, it is shown that the proposal decreases the number of unnecessary handovers by 70\%.

\subsection{Lessons Learned}

First, based on \cite{M-Conf-2,M-Conf-3,M-4,M-Conf-4}, fuzzy Q learning is a common approach to
handover parameter optimization, and KPIs, such as
radio link failure, handover failure and CDR,
are suitable candidates for the observation state because of their close relationship with the handover process.
Second, according to \cite{1,M-Arxiv-1}, deep reinforcement learning can be used to directly
make handover decisions on the user-BS association taking only the measurements from users like RSSI and RSRQ
as the input.
Third, following \cite{M-2}, the lack of user history trajectory information can be overcome by
first predicting the user's next activity and then predicting its next location.
At last, as indicated by \cite{M-7}, unsupervised-shapelets can help
find useful patterns from RSRP information reported by users and further make
BSs aware of the RF conditions at their cell edge, which is critical to
handover parameter setting.

\begin{table*}[!t]
\caption{Machine Learning Based Mobility Management}
\scriptsize
\centering
\begin{tabular}{p{1cm}p{2cm}p{4 cm}p{2 cm}p{6 cm}}
\toprule
\textbf{Literature} & \textbf{Scenario}  & \textbf{Objective} & \textbf{Machine learning methods} & \textbf{Main conclusion}\\
\midrule 
\cite{M-Conf-2}  & A small cell network & Achieve a tradeoff between the signaling cost led by handover and the user experience influenced by CDR & Fuzzy Q learning & The proposal can minimize the number of handovers while keeping the CDR at a desired level\\
\midrule

\cite{M-Conf-3}  & A cellular network & Minimize the weighted difference between the target KPIs and achieved KPIs & Fuzzy Q learning & The proposal outperforms existing methods, in terms of HO failures and ping-pong HOs\\
\midrule

\cite{M-4}  & A small cell network & Achieve load balancing & Fuzzy Q learning & The proposal can minimize the localized congestion of small-cell networks\\
\midrule

\cite{M-Conf-4}  & An LTE network & Realize the coordination among multiple SON functions & Fuzzy Q learning & The proposal can enable the joint optimization of different KPIs \\
\midrule

\cite{1}  & A WLAN & Optimize system throughput & Deep reinforcement learning & The proposal can improve system throughput and reduce the number of handovers \\
\midrule

\cite{M-Arxiv-1}  & A cellular network & Minimize handover rate and ensure system throughput & Deep reinforcement learning & The proposal outperforms the
state-of-art on-line schemes in terms of HO rate \\
\midrule

\cite{M-2}  & A wireless network & Predict users' next location & Supervised learning & The proposed approach can predict more accurately and perform robustly \\
\midrule

\cite{M-7}  & A heterogeneous network  & Classify the user trajectories  & Unsupervised shapelets & The proposed approach provides clustering results with an average accuracy of 95\% \\
\midrule

\cite{M-8}  & An LTE network  & Identify whether an indoor user should be switched to an external BS & Self-organizing map & The proposed approach can reduce unnecessary handovers by up to 70\% \\
\bottomrule
\end{tabular}
\end{table*}

\section{Machine Learning Based Localization}  

In recent years, we have witnessed an explosive proliferation of location based services, whose service quality is highly dependent on the accuracy of localization.
The mature technique Global Positioning System (GPS) has been widely used for outdoor localization.
However, when it comes to indoor localization, GPS signals from a satellite will be heavily attenuated, which makes GPS incapable of use in indoor localization.
Furthermore, indoor environments are more complex as there are lots of obstacles such as tables, wardrobes, and so on, thus causing the difficulty of localization.
In this situation, to locate indoor mobile users precisely, many wireless technologies can be utilized such as WLAN, ultra-wide bandwidth (UWB) and Bluetooth.
Moreover, common measurements used in indoor localization include time of arrival (TOA), TDOA,
channel state information (CSI) and received signal strength (RSS).
To solve various problems associated with indoor localization,
research has been conducted by adopting machine learning in scenarios with different wireless technologies.

\subsubsection{Supervised Learning Based Approaches}

Instead of assuming that equal signal differences account
for equal geometrical distances as in traditional KNN based localization approaches do,
authors in \cite{ind20} propose a feature scaling based KNN (FS-KNN) localization algorithm.
This algorithm is inspired by the fact that the relation between signal differences and geometrical distances is actually dependent on the measured RSS.
Specifically, in the signal distance calculation between the fingerprint of each RP and the RSS vector reported by the user,
the square of each element-wise difference is multiplied by a weight that is a function of the corresponding RSS value measured by the user.
To identify the parameters involved in the weight function, an iterative training procedure is used, and
performance evaluation on a test set is made, whose metric is taken as the objective function of a simulated annealing algorithm to
tune those parameters.
After the model is well trained, the distance between a newly received RSS vector and each fingerprint is calculated, and then
the location of the user is determined by calculating a weighted mean of the locations of the k nearest RPs.

To deal with the high energy consumption incurred by frequent AP scanning via WiFi interfaces,
an energy-efficient indoor localization system is developed in \cite{ind10}, where
ZigBee interfaces are used to collect WiFi signals. To improve localization accuracy,
three KNN based localization approaches adopting different distance metrics are evaluated, including
weighted Euclidian distance, weighted Manhattan distance and relative entropy.
The principle for weight setting is that the AP with more redundant information is assigned with a lower weight.
In \cite{ind21}, authors theoretically analyze the optimal number of nearest RPs used to identify the user location in a KNN based localization algorithm,
and it is shown that $k=1$ and $k=2$ outperform the other settings for static localization.

To avoid regularly training localization models from scratch, authors
in \cite{ind1} propose an online independent support vector machine (OISVM) based localization system
that employs the RSS of Wi-Fi signals.
Compared to traditional SVM, OISVM is capable of learning in an online fashion and allows to make a balance between accuracy and model size,
facilitating its adoption on mobile devices.
The constructed system includes two phases, i.e., the offline phase and the online phase.
The offline phase further includes kernel parameter selection, data under sampling to deal with the imbalanced data problem,
and offline training using pre-collected RSS data set with RSS samples appended with corresponding RP labels.
In the online phase, location estimation is conducted for new RSS samples, and meanwhile
online learning is performed as new training data arrives, which can be collected via crowdsourcing.
The simulation result shows that the proposal can reduce the location estimation error by 0.8m, while the prediction time and training time
are decreased significantly compared to traditional methods.

Given that non-line-of-sight (NLOS) radio blockage can lower the localization accuracy,
it is beneficial to identify NLOS signals.
To this end, authors in \cite{ind7} develop a relevance vector machine (RVM) based method
for ultrawide bandwidth TOA localization.
Specifically, a RVM based classifier is used to identify the LOS and NLOS signals received by
the agent with unknown position from the anchor whose position is already known,
while a RVM regressor is adopted for ranging error prediction.
Both of the two models take a feature vector as input data, which consists of received energy,
maximum amplitude, and so on.
The advantage of RVM over SVM is that RVM uses a smaller number of
relevance vectors than the number of support vectors in the SVM case, hence reducing the computational complexity.
On the contrary, authors in \cite{ind2} propose an SVM based method,
where a mapping between features extracted from the received
waveform and the ranging error is directly learned.
Hence, explicit LOS and NLOS signal identification is not needed anymore.

In addition to KNN and SVM, some researchers have applied NNs to localization.
In order to reduce the time cost of the training procedure, authors in \cite{ind16} utilize ELM.
The RSS fingerprints and their corresponding physical coordinates are used to train the output weights.
After the model is trained, it can predict the physical coordinate for a new RSS vector.
Also adopting a single layer NN, in \cite{ind31}, an NN based method is proposed for
an LTE downlink system, aiming at estimating the UE position.
The employed NN contains three layers, namely an input layer, a hidden layer and an output layer, with the input and the output being channel parameters and the corresponding coordinates, respectively.
Levenberg Marquardt algorithm is applied to iteratively adjust the weights of the NN based on the Mean Squared Error that is a
metric to assess the error between the output and the corresponding known label. When the NN is trained, it can predict the location given the new data.
Preliminary experimental results show that the proposed method yields a median positioning error distance of 6 meters for the indoor scenario.

In a WiFi network with RSS based measurement, a deep NN is utilized in \cite{ind32} for indoor localization without using the pathloss model or
comparing with the fingerprint database.
The training set contains the RSS vectors appended with the central coordinates and indexes of the corresponding grid areas.
The implementation procedure of the NN
can be divided into three parts that are transforming, denoising and localization.
Particularly, this method pre-trains the transforming section and denoising section by using auto-encoder block.
Experiments show that the proposed method can realize higher localization accuracy compared with
maximum likelihood estimation, the generalised regression
neural network and fingerprinting methods.

\subsubsection{Unsupervised Learning Based Approaches}

To reduce computation complexity and save storage space for fingerprinting based localization systems,
authors in \cite{ind19} first divide the radio map into multiple sub radio maps,
and then use Kernel Principal Components Analysis (KPCA) to extract features of each sub radio map
to get a low dimensional version.
Result shows that the size of the radio map can be reduced by 72\% with 2m localization error.
In \cite{ind23}, PCA is also employed together with linear discriminant analysis
to extract lower dimensional features from raw RSS measurement.

Considering the drawbacks of adopting RSS measurement as fingerprints like its high randomness
and loose correlation with propagation distance,
authors in \cite{ind4} propose to utilize the calibrated CSI phase information
for indoor fingerprinting.
Specifically, in the off-line phase, a deep autoencoder network is constructed for each position to
reconstruct the collected calibrated CSI phase information,
and the weights are recorded as the fingerprints.
In the online phase, a new location is obtained by a probabilistic method that performs a weighted average of all
the reference locations.
Simulation results show that
the proposal outperforms traditional CSI or RSS based methods in two representative scenarios.


Moreover, ideas similar to \cite{ind4} are adopted in \cite{xvyu} and \cite{xvyu1}.
In \cite{xvyu}, CSI amplitude responses are taken as the input of the deep NN, which
is trained by a greedy learning algorithm layer by layer to reduce complexity.
After the deep NN is trained, the weights in the deep network are stored
as fingerprints to facilitate localization in the online test.
Instead of using only CSI phase information or amplitude information,
Wang et al. in \cite{xvyu1} propose to use a deep autoencoder network to extract channel features from
bi-modal CSI data, containing average amplitudes and estimated angle of arrivings,
and the weights of the deep autoencoder network are seen as the extracted features (i.e., fingerprints).
Owing to bi-modal CSI data, the proposed approach achieves higher localization precision than
several representative schemes in literatures.
In \cite{3d}, authors introduce a
denoising autoencoder for bluetooth, low-energy, indoor localization to provide high performance 3-D localization
in large indoor places.
Here, useful fingerprint patterns hidden in the received signal strength indicator measurements are extracted by the autoencoder,
which helps the construction of the fingerprint database where the reference locations are in 3-D space.
By field experiments, it is shown that 3-D space fingerprinting contributes to higher
positioning accuracy, and the proposal outperforms comparison schemes, in terms of horizontal accuracy and vertical accuracy.

\subsubsection{Lessons learned}

First, from \cite{ind20,ind10}, it can be learned that the most intuitive approach to indoor localization
is to use KNN that depends on the similarity between the RSS vector measured by the user and
pre-collected fingerprints at different RPs.
Second, as indicated by \cite{ind20},
the relation between signal differences and geometrical distances relies on the measured RSS.
Third, following \cite{ind21}, for a KNN based localization algorithm applied to static localization,
$k=1$ and $k=2$ are better choices than other settings of
the number of the nearest RPs used to identify the user location.
Fourth, based on \cite{ind7}, the RVM classifier is preferred
for the identification of LOS and NLOS signals compared to the SVM classifier,
since the former has lower computation complexity owing to less number of relevance vectors.
Moreover, according to \cite{ind19}, the size of the radio map can be effectively reduced by
KPCA, saving the storage capacity of user terminals.
At last, as revealed in \cite{3d,ind4,xvyu,xvyu1},
autoencoder is able to extract useful and robust information from RSS data or CSI data,
which contributes to higher localization accuracy.

\begin{table*}[!t]
\caption{Machine Learning Based Localization}
\scriptsize
\centering
\begin{tabular}{p{1cm}p{2cm}p{4 cm}p{2 cm}p{6 cm}}
\hline
\textbf{Literature} & \textbf{Scenario}  & \textbf{Objective} & \textbf{Machine learning methods} & \textbf{Main conclusion}\\
\midrule 
\cite{ind20}  & WLAN & Improve localization accuracy by considering a more reasonable similarity metric for RSS vectors & FS-KNN & Simulation shows that FS-KNN can achieve an average distance error of 1.93m\\
\midrule
\cite{ind10}  &WLAN & Improve the energy efficiency of indoor localization & KNN & The proposal can achieve high localization accuracy with the average energy consumption significantly reduced compared with WiFi interface based methods\\
\midrule
\cite{ind21}  &WLAN& Find the optimal number of nearest RPs for KNN based localization& KNN & The accuracy performance can
not be improved by further increasing the optimal parameter k \\
\midrule

\cite{ind1}  & WLAN & Reduce training and prediction time & Online independent support vector machine & The proposal can improve localization accuracy and meanwhile decrease the prediction time and training time \\
\midrule

\cite{ind7}  & UWB & Identify NLOS signal & Relevance vector machine  & The simulation result shows that the proposal can improve the range estimates of NLOS signals\\
\midrule

\cite{ind2}  & UWB & Mitigate ranging error without NLOS signal identification & SVM &  The proposal can achieve considerable performance improvements in various practical
localization scenarios\\
\midrule

\cite{ind19} & WLAN & Reduce the size of radio map & KPCA & The proposal can reach high localization accuracy while reducing 74\% size of the radio map\\
\midrule

\cite{ind23}  &WLAN& Extract low dimensional features of RSS data & PCA and LCA &The combination method outperforms the LCA-based method in terms of the accuracy of floor classification \\
\midrule


\cite{ind16}  &WLAN & Reduce localization model training time & Extreme learning machine & The consensus-based parallel ELM performs competitively compared to
 centralized ELM, in terms of localization accuracy, with more robustness and no additional computation cost \\
\midrule

\cite{ind31}  & An LTE network & Reduce calculation time & Dense neural network & By using only one LTE eNodeB, the proposal can achieve an error distance of 6 meters in indoor environments\\
\midrule

\cite{ind32}  &WLAN & Achieve high localization accuracy without using the radio pathloss model or comparing with the radio map  & Auto-encoder & The proposal outperforms maximum likelihood estimation, the generalised regression neural network and fingerprinting methods\\
\midrule

\cite{ind4}  &WLAN & Achieve higher localization accuracy by utilizing CSI phase information for fingerprinting & Deep autoencoder network  & The proposal outperforms three benchmark schemes based on CSI or RSS in two typical indoor scenarios\\
\midrule

\cite{xvyu}  &WLAN & Overcome the drawbacks of RSS based localization methods by utilizing CSI information for fingerprinting & Deep autoencoder network & Experimental results show
that the proposal can localize the target effectively\\
\midrule

\cite{xvyu1}  &WLAN & Achieve higher localization accuracy by utilizing CSI amplitudes and estimated angle of arrivals for fingerprinting & Deep autoencoder network  &
The proposal has superior performance compared to several baseline schemes like FIFS in \cite{duibi}\\
\midrule

\cite{3d}  &WLAN & Achieve high performance 3-D localization in large indoor places & Deep autoencoder network  &
 Positioning accuracy can be effectively improved by 3-D space fingerprinting  \\
\bottomrule
\end{tabular}
\end{table*}

\section{Conditions for The Application of Machine Learning}

In this section, several conditions for the application of ML are elaborated, in terms of
the type of the problem to be solved, training data,
time cost, implementation complexity and differences between machine learning techniques in the same category.
These conditions should be checked one-by-one before making the final decision about whether to adopt ML techniques
and which kind of ML techniques to use.

\subsection{The Type of The Problem}

The first condition to be considered is the type of the concerned wireless communication problem.
Generally speaking, problems addressed by machine learning can be categorized into regression problems,
classification problems, clustering problems and Markov decision making problems.
In regression problems, the task of machine learning is to predict continuous values for the current input,
while machine learning should identify the class to which the current input belongs in classification problems.
As for Markov decision making problems, machine learning needs to output a policy that guides
the action selection under each possible system state.
In addition, all these kind of problems may involve a procedure of feature extraction that can be done manually or
in an algorithmic way.
If the studied problem lies in one of these categories, one can take machine learning as a possible solution.
For example, content caching may need to acquire content request probabilities of each user.
This problem can be seen as a regression problem that takes the user profile as input and the content request probabilities as output.
For BS clustering, it can be naturally handled by K-means clustering algorithms, while many resource management
and mobility management problems introduced in this paper are modeled as a Markov decision making problem that
can be efficiently solved by reinforcement learning.
In \cite{knnbeam}, the beam allocation problem is transformed into a classification problem that can be easily addressed by KNN.
\emph{In summary, the first condition for applying machine learning is that the considered wireless communication problem can be
abstracted into tasks that are suitable for machine learning to handle.}

\subsection{Training Data Availability}

In surveyed works, training data for supervised and unsupervised learning
is generated or collected in different ways depending on specific problems.
Specifically, for power allocation, the aim of authors in
\cite{nnbeam,nnpower,auto-power} is to utility neural networks
to approximate high complexity power allocation algorithms, such as the genetic algorithm and
WMMSE algorithm. At this time, the training data is generated by run these algorithms under
different network scenarios for multiple times.
In \cite{3,vrchen}, since spectrum allocation among BSs is modeled as a non-cooperative game,
the data to train the RNN model at each BS is generated by the continuous interactions
of BSs. For cache management, authors in \cite{tonycache} utilize
a mixture of data set YUPENN \cite{data1} and UFC101 \cite{data1}
as their own data set, while authors in \cite{elmcache}
collect their data set by using the YouTube Application Programming
Interface, which consists of 12500 YouTube videos.
In \cite{chenjsac,chencran}, the content request data that the RNN uses
to train and predict content request distribution is obtained
from Youku of China network video index.

In \cite{knnbeam} studying KNN based beamformer design,
a feature vector in the training data set under a specific resource allocation scenario is composed of
time-variant parameters, such as the number of users and CSI,
and these parameters can be collected and stored by the cloud.
In \cite{S-20}, DNN based routing is investigated, and
the training data set is obtained by
recording the traffic patterns and paths while running traditional routing
strategies such as OSPF.
In \cite{S-29}, the training data set is generated in an online
fashion, which is collected by each router in the network.
In literatures \cite{3d,ind20,ind10,ind21,ind1,ind16,ind31,ind32,ind4,xvyu,xvyu1}
focusing on machine learning based localization,
training data is based on CSI data, RSSI data or some channel parameters
and comes from practical measurement in a real scenario using a certain device like a cell phone.
To obtain these wireless data, different ways can be adopted.
For example, authors in \cite{ind4} uses one mobile device equipped with an
IWL 5300 NIC that can read CSI data from the slight modified device driver,
while authors in \cite{ind20} develop a client program for the mobile device to
facilitate RSS measurement.

As for other surveyed works, they mainly adopt reinforcement learning
to solve resource management, networking and mobility management problems.
In reinforcement learning, the reward/cost, which is fed back by the environment
after the learning agent takes an action, can be seen as training data.
The environment is often an virtual environment created by using certain softwares like Matlab, and
the reward function is defined to reflect the objective of the
studied problem.
For example, authors in \cite{sun} aim at optimizing system spectral efficiency, which is hence
taken as the reward fed back to each D2D pair, while the reward for the DRL agent in \cite{tangjian}
is defined as the difference between the maximum possible total power consumption
and the actual total power consumption, which helps minimize system power cost.
\emph{In summary, the second condition for the application of ML is that
essential training data can be acquired.}

\subsection{Time Cost}

Another important aspect, which can prevent the application of machine learning,
is the time cost.
Here, it is necessary to distinguish between two different time metrics, namely
training time and response time as per \cite{son}.
The former represents the amount of time that ensures a machine learning
algorithm is fully trained, which is important for supervised and unsupervised learning
to make accurate predictions for future inputs and also important for reinforcement learning
to learn a good strategy or policy.
As for response time, for a trained supervised or unsupervised learning algorithm,
response time refers to the time needed to output a prediction given an input,
while it refers to the time needed to output an action for a trained reinforcement learning model.

In some applications,
there can be a stringent requirement about response time.
For example, resource management decisions should be made
on a timescale of milliseconds.
To figure out the feasibility of machine learning in these applications,
we first make a discussion from the perspective of response time.
In our surveyed papers, machine learning techniques applied in resource management
can be coarsely divided into neural network based approaches and
other approaches.

1)\emph{Neural network based approaches:}
For the response time of neural networks after trained, it has been reported that
Graphics Processing Unit (GPU)-based parallel computing can enable them to make
predictions within milliseconds \cite{dlbook}.
Note that even without GPUs, a trained DNN can make a
power control decision for a network with 30 users using 0.0149 ms on average (see Table I in \cite{nnbeam}).
Hence, it is possible for the proposed neural network based power allocation approaches in
\cite{nnbeam,nnpower,auto-power} to make resource management decisions in time.
Similarly, since deep reinforcement learning selects the
resource allocation action based on the Q-values
output by the deep Q network, which is actually a neural network,
a trained deep reinforcement learning model can also be suitable for resource management in wireless networks.

2)\emph{Other approaches:}
First, traditional reinforcement learning, which includes Q learning
and joint utility and strategy estimation based learning,
aims at deriving a strategy or policy for the learning agent under a dynamic environment.
Once reinforcement learning algorithms converge after being trained for sufficient time,
the strategy or policy becomes fixed.
In Q learning, the policy is represented by a set of Q values, each of which associates with
a pair of a system state and an action, and the learning agent chooses the
resource allocation action with the maximal Q value under the current system state.
As for the joint utility and strategy estimation based learning,
the agent's strategy is composed of probabilities to select each action, and the agent
only needs to generate a random number between 0-1 to identify which action to play.
Therefore, a well-trained agent using these two learning techniques can make
quick decisions within milliseconds.
Second, for the KNN based resource management adopted in \cite{knnbeam}, it relies on the calculation of
similarity between the feature vector of the new network scenario and that of each past network scenario.
However, owing to cloud computing, the similarities can be calculated in parallel.
Hence, the most similar $K$ past scenarios can be identified within very short time,
and the resource management decision can be made very quickly by directly taking the
resource configuration adopted by the majority of the $K$ past scenarios.

Next, the discussion is made from the perspective of training time,
and note that the KNN based approach does not have a training procedure.
Specifically,
for a deep NN model,
its training may take a long time. Therefore, it is possible that
the patterns of the communication environment has changed before the model learns a mapping rule.
Moreover, training a reinforcement learning
model in a complex environment can also cost much time, and it is possible that
the elements of the communication environment, such as the environmental dynamics and the set of agents, have been different before training is completed.
Such a mismatch between training time and the timescale on which the characteristics of communication environments change
can do harm to the actual performance of a trained model.
Nevertheless, training time can be reduced for neural network based approaches with the help of GPUs and transfer learning,
while the training of traditional reinforcement learning can be accelerated using transfer learning as well, as shown in \cite{bs3}.
\emph{In summary, the third condition for the application of machine learning is that time cost, including response time and training time,
should meet the requirements of applications and match with the time scale on which the communication environment varies.
}

\subsection{Implementation Complexity}

In this subsection, the implementation complexity of machine learning algorithms is discussed by
taking the algorithms for mobility management as examples.
Related surveyed works can be categorized into directly optimizing handover parameters
using fuzzy Q learning \cite{M-Conf-2,M-Conf-3,M-4,M-Conf-4}, handing over users to proper BSs using deep reinforcement learning \cite{1,M-Arxiv-1},
and helping improve handover performance by predicting user's next location using a probabilistic model \cite{M-2}, analyzing RSRP patterns using unsupervised shapelets \cite{M-7}
and identifying the area where handover is prohibited using self-organizing map\cite{M-8}.

First, to implement fuzzy Q learning, a table needs to be maintained to store q values, each of which
corresponds to a pair of a rule and one of its actions. In addition, only simple mathematical operations
are involved in the learning process, and the inputs are common network KPIs like call dropping ratio
or the value of the parameter to be adjusted, which can all be easily acquired by the current network.
Hence, it can be claimed that fuzzy Q learning possesses
low implementation complexity.
Second, deep reinforcement learning is based on neural networks, and it
can be conveniently implemented by utilizing rich frameworks for deep learning, such as TensorFlow
and Keras. Its inputs are Received Signal Strength Indicator (RSSI) in \cite{1} and
Reference Signal Received Quality (RSRQ) in \cite{M-Arxiv-1}, which can both be collected by the current system.
However, to accelerate the training process, it is preferred to
run deep reinforcement learning programs on GPUs, and this may incur high cost for the implementation.
Third, the approach in \cite{M-2} is based on a simple probabilistic model, while the approaches
in \cite{M-7,M-8} only utilize the information reported by user terminals in current networks.
Hence, the proposals in \cite{M-2,M-7,M-8} can be not that complex for practical implementation.

Although only the implementation complexity of machine learning methods for mobility management
is discussed, other surveyed methods can be analyzed in a similar way.
Specifically, the implementation complexity should consider the complexity of data storage,
the mathematical operations involved in the algorithms,
the complexity of collecting the necessary information and the requirement on softwares and hardwares.
\emph{In summary, the fourth condition for the application of machine learning is that implementation complexity
should be acceptable.}

\subsection{Comparison Between Machine Learning Techniques}

Problems in surveyed works can be generally
categorized into regression, classification, clustering and decision making.
However, for each kind of problem, different machine learning techniques can be available.
In this section, comparison between machine learning methods that can handle problems of the same type is conducted, which
reveals the reason for surveyed works to adopt a certain machine learning technique instead of others.
Most importantly, readers can be guided to select the suitable machine learning technique.

1)\emph{Machine Learning Techniques for Regression and Classification:}
Machine learning models applied to regression
and classification tasks in surveyed works mainly include SVM, KNN and neural networks.
KNN is a basic classification algorithm that is known to be very
simple to implement.
Generally, KNN is used as multi-class classifiers whereas
standard SVM has been regarded as one of the
most robust and successful algorithms to design
low-complexity binary classifiers \cite{knnbeam}.
When data is not linearly separable, KNN can be a good choice compared to SVM.
This is because the regularization term and the kernel parameters should be selected for SVM,
while one needs to choose only the k parameter and the distance metric for KNN.
Compared with SVM and KNN, deep neural networks are powerful in feature extraction and the performance
can be improved more significantly with the increase of training data size.
However, due to the optimization of a large number of parameters, their training
can be time consuming.
Hence, when enough training data and GPU resource are available, deep neural networks are preferred.
In addition, common neural networks further include DNN, CNN, RNN and extreme learning machine.
Compared with DNN, the number of weights can be reduced by CNN, which makes the training and inference procedures
faster with lower overheads. In addition, CNN is good at learning spatial features, such as the features
of a channel gain matrix.
RNN is suitable for processing time series to learn features in time domain, while
the advantage of extreme learning machine lies in
good generalization performance at an extremely fast
learning speed without iteratively tuning on the hidden layer
parameters.

2)\emph{Machine Learning Techniques for Decision Making:}
Machine learning algorithms applied to decision making in dynamic environments
in surveyed works mainly include actor-critic learning, Q learning,
joint utility and strategy estimation
based learning (JUSEL) and deep reinforcement learning.
Compared with Q learning, actor-critic learning is
able to learn an explicit stochastic policy
that may be useful in non-Markov environments \cite{ac}.
In addition, since value function and policy are updated separately,
policy knowledge transfer is easier to achieve \cite{kt}.
For JUSEL, it is very suitable in multi-agent scenarios
and is able to achieve stable systems where
the gain obtained is bounded when an agent unilaterally deviates from
its mixed strategy.
Compared with Q learning and actor-critic learning, one of the advantages of deep reinforcement learning
lies in its ability to learn from high dimensional input states, owing to the deep Q network.
On the contrary, since both Q learning and actor-critic learning need to store an evaluation for each state-action pair,
they are not suitable for communication systems with states of large dimension.
Another advantage of deep reinforcement learning is its ability to infer a good action under an unfamiliar state \cite{yuyi1}.
Nevertheless, training deep reinforcement learning can incur high computing burden.
Finally, in addition to deep reinforcement learning, fuzzy Q learning, as a variant of Q learning, can also
address the situation of continuous states but with lower computation cost.
However, setting membership functions requires prior experience, and the number of rules in the rule base
can exponentially increase when the state dimension is high.

\emph{In summary, the fifth condition for the application of machine learning is that
the advantages of the adopted machine learning technique well fit into the studied problem, and meanwhile
its disadvantages are tolerable.}

\section{Alternatives to Machine Learning and Motivations}

In this section, we review and elaborate traditional approaches
that are taken as baselines in the surveyed works and are not based on machine learning.
By comparing with these alternatives, the motivations to adopt machine learning
are carefully summarized.

\subsection{Alternatives for Power Allocation}

Basic alternatives are some simple heuristic
schemes, such as uniform power allocation among resource blocks \cite{24}, transmitting with full power \cite{42}
and smart power control \cite{Add-1}. However, considering the sophisticated radio environment faced by
each BS, where the power allocation decisions are coupled among BSs, these heuristic schemes can
have poor performance. Specifically, in \cite{24}, it is reported that
the proposed Q learning based scheme achieves 125\% performance improvement
compared to uniform power allocation, while the average femtocell capacity
is enhanced by 50.79\% using Q learning compared with smart power control in \cite{Add-1}.
When power levels are discrete, numerical search can be adopted, including
exhaustive search and genetic algorithm that
is a heuristic searching algorithm inspired by the theory of natural
evolution \cite{ga}.
In \cite{sun13}, it is shown that multi-agent Q learning can reach near optimal
performance but with a huge reduction in control signalling compared to
centralized exhaustive search. In \cite{auto-power}, the trained deep learning model
based on auto-encoder is capable of outputting the
same resource allocation solution got by genetic algorithm 86.3\% of time
with less computation complexity.
Another classical approach to power allocation is the WMMSE algorithm utilized
to generate the training data in \cite{nnbeam,nnpower}.
The WMMSE algorithm is originally designed to optimize beamformer
vectors, which transforms the weighted sum-rate maximization problem into a higher
dimensional space to make the problem more tractable \cite{wmmse}.
In \cite{nnbeam}, for a network with 20 users, the DNN based power control algorithm is demonstrated to
achieve over 90\% sum rate got by WMMSE algorithm, but its CPU
time only accounts for 4.8\% of the latter's CPU time.

\subsection{Alternatives for Spectrum Management}

In \cite{sun40}, the proposed reinforcement learning based spectrum management scheme
is compared with a centralized dynamic spectrum sharing (CDSS) scheme developed in \cite{cdss}.
The simulation result reveals that reinforcement learning
can reach nearly the same average cell throughput
as the CDSS approach without information sharing between BSs.
In \cite{sun}, the adopted distributed reinforcement learning is shown to
achieve similar system spectral efficiency got by exhaustive search that
can bring a huge computing burden on the cloud.
Moreover, a simple way of resource block allocation is the
proportional fair algorithm \cite{pro} that is utilized as a baseline
in \cite{vrchen}, where the proposed RNN based resource allocation approach
significantly outperforms the proportional fair algorithm, in terms of user delay.

\subsection{Alternatives for Backhaul Management}

In \cite{Conf-36}, Q learning based cell range extension offset (CREO) adjustment
greatly reduces the number of users in cells with congested backhaul compared to
a static CREO setting.
In \cite{75}, a branch-and-bound based centralized approach is developed as
the benchmark, aiming at maximizing the total backhaul resource utilization.
Compared to the benchmark,
the proposed distributed Q learning can achieve a competitive performance, in
terms of the average throughput per user.
Authors in \cite{Backhaul+} use a centralized greedy backhaul management strategy as a comparison.
The idea is to identify some BSs to download a fixed number of predicted files
based on a fairness rule. According to the simulation,
reinforcement learning based backhaul management can
reach much higher performance
than centralized greedy approach, in terms of remaining backhaul capacity, at a lower signalling cost.
In \cite{46}, to verify the effectiveness of the reinforcement learning based method,
a simple baseline is that the messages of MUEs are not transmitted via backhaul links
between SBSs and the MBS, which leads to poor MUE rates.

\subsection{Alternatives for Cache Management}

In \cite{benecache}, random caching and caching based on time-averaged content popularity are taken as
baselines, and reinforcement learning based cache management achieves 13\% and 56\% higher
per-BS utility than these two heuristic schemes for a dense network scenario.
Also compared with the two schemes, the extreme learning machine based caching scheme in \cite{elmcache}
decreases downloading delay significantly.
Moreover, the transfer learning based caching in \cite{tlcache2}
outperforms random caching and achieves close performance to caching the most popular contents
with popularity perfectly known, in terms of backhaul load and user satisfaction ratio.
Another two well-known heuristic cache management strategies are
least recently used (LRU) strategy \cite{lru} and least frequently used (LFU) strategy \cite{lfu}.
Under LRU strategy, the cached content, which is least requested recently, will be
replaced by the new content when the cache storage is full, while
the cached content, which is requested the least many times, will be replaced under LFU strategy.
In \cite{weicache}, Q learning is shown to greatly outperform LRU and LFU strategies in reducing
transmission cost, while the deep reinforcement learning approach in \cite{drlcache} achieves
higher cache hit rate compared with these two strategies.

\subsection{Alternatives for Beamforming}

In \cite{knnbeam}, authors use KNN to address a beam allocation problem.
The alternatives include exhaustive search and a low complexity beam allocation (LBA) algorithm
proposed in \cite{beam}. The latter is based on submodular optimization theory that
is a powerful tool for solving combinatorial optimization problems.
Via simulation, it is observed that the
KNN based allocation algorithm can approach optimal average sum rate
and outperforms LBA algorithm with the increase of training data size.

\subsection{Alternatives for Computation Resource Management}

In \cite{chenxian}, three heuristic computation offloading mechanisms are taken as baselines,
namely mobile execution, server execution and greedy execution.
In the first and second schemes, the mobile user processes computation tasks locally
and offloads computation tasks to the MEC server, respectively.
For the greedy execution, the mobile user makes the offloading decision
to minimize the immediate task execution delay.
Numerical results reveal that
much better long-term utility performance can be achieved by deep reinforcement
learning based approach compared with these baselines, and meanwhile
the proposal does not need to know the information of network dynamics.

\subsection{Alternatives for User Association}

In \cite{ha}, it is observed that multi-armed bandits learning based association can
achieve similar performance to exhaustive search but with
lower complexity and lower overhead caused by information acquisition.
In \cite{U-1}, reinforcement learning based user association is compared with
two common association schemes. One is max-SINR based association,
which means each user chooses the BS with the maximal SINR to associate, and the other one
is based on optimization theory, which adopts gradient descent and dual decomposition \cite{3d1}.
By simulation, it can be seen that reinforcement learning can
make user experience more uniform and meanwhile deliver higher
rates for vehicles. The max-SINR based association is also used as a baseline in
\cite{U-3}, which leads to poor QoS of UEs.

\subsection{Alternatives for BS Switching Control}

Assuming a full knowledge of traffic loads,
authors in \cite{prior} greedily turn off as many as BSs
to get the optimal BS switching solution,
which is taken by \cite{bs3} as a comparison scheme to verify
the effectiveness of transfer learning based BS switching.
In \cite{bs2}, it is observed that actor-critic based BS switching consumes only a little more
energy than an exhaustive search based scheme.
However, the learning based approach does not need the knowledge
of traffic loads in advance.
In \cite{tangjian}, two alternative schemes to control BS on-off states are considered,
namely single-BS association (SA) and full coordinated association (FA).
In SA scheme, each user associates with a BS randomly and BSs without serving any users
are turned off, while all the BSs are active in FA scheme.
Compared to the heuristic approaches, deep reinforcement learning based method
can achieve lower energy consumption while meeting users' demands.

\subsection{Alternatives for Network Routing}

In \cite{S-21}, authors utilize spectrum-aware ad hoc on-demand distance vector routing (SA-AODV) approach
as a baseline to demonstrate the superiority of their reinforcement learning based routing.
Specifically, their proposal leads to lower route discovery frequency.
In \cite{S-31}, two intuitive routing strategies are presented for performance comparison, which
are shortest path (SP) routing and optimal primary user (PU) aware shortest path
(PASP) routing. The first scheme aims at minimizing the number of hops in the route,
while the second scheme intends to minimize the accumulated amount of PUs' activities.
It has been shown that the proposed learning based routing can cause lower
end-to-end delay than the two schemes.
According to Fig. 4 in \cite{S-20}, the deep learning based routing algorithm
reduces average per-hop delay by around 93\% compared with
open shortest path first (OSPF) routing.
In \cite{S-29}, OSPF is also taken as a baseline that always leads to high delay and packet loss
rate. Instead, the proposed deep convolutional neural network based routing strategy
can reduce them significantly.

\subsection{Alternatives for Clustering}

In \cite{yufei}, deep reinforcement learning is used to select transmitter-receiver pairs
to form a cluster, within which cooperative interference alignment is performed.
By comparing with two existing user selection schemes proposed in \cite{us1} and \cite{us2}
that are designed for static environments, deep reinforcement learning is shown to be capable of
greatly improving user sum rate in the environment with time-varying CSI.

\subsection{Alternatives for Mobility Management}

In \cite{1}, authors compare their deep reinforcement learning approach with a handover policy
in \cite{hp} that compares the RSSI values from the current AP and other APs.
Simulation result indicates that deep reinforcement learning can mitigate
ping-pong effect with high data rate.
In \cite{M-Conf-3}, a fuzzy Q learning algorithm is developed to adjust hysteresis and time-to-trigger values.
To verify the effectiveness of the algorithm, two baseline schemes are considered, namely
trend-based handover optimization proposed in \cite{tb} and a scheme setting time-to-trigger values based
on velocity estimates.
As for performance comparison, it is observed that fuzzy Q learning based hysteresis adjustment
significantly outperforms the two baselines, in terms of the number of early handover.
Another alternative for mobility management is using fuzzy logic controller (FLC).
In \cite{M-Conf-2} and \cite{M-4}, numeric simulation has demonstrated the advantages of fuzzy Q learning over
FLC whose performance is limited by available prior knowledge.
Specifically, it is reported in \cite{M-Conf-2} that
fuzzy Q learning can still achieve competitive performance even without enough prior knowledge,
while it is is shown to reach better long-term performance in \cite{M-4}
compared with the FLC based method.

\subsection{Alternatives for Localization}

Surveyed works mainly utilize localization methods that are based on probability theory for performance comparison.
These methods include FIFS \cite{duibi} and Horus \cite{horus}.
In \cite{ind20}, the simulation result shows that the mean of error distance achieved by the proposed feature scaling based KNN localization algorithm
is 1.82 times better than that achieved by Horus.
In \cite{xvyu}, the deep auto-encoder based approach improves the mean of the location errors
by 20\% and 31\% compared to FIFS and Horus, respectively.
The superiority of deep learning in enhancing localization performance has also been
verified in \cite{ind4,xvyu1}.
In \cite{ind2}, authors propose to use machine learning to
to estimate the ranging error for UWB localization, and they make comparisons with
two schemes purely based on norm optimization without ranging error mitigation,
which leads to poor localization performance.
The other surveyed papers like \cite{ind16} and \cite{ind32} mainly compare their proposals with
approaches that are also based on machine learning.

\subsection{Motivations to Apply Machine Learning}

After summarizing traditional schemes, the motivations of authors in surveyed literatures to
adopt machine learning based approaches are clarified as follows.
\begin{itemize}
  \item \emph{Developing Low-complexity Algorithms for Wireless Problems:} This is a main reason for researchers to use deep neural networks to approximate high complexity resource allocation algorithms. Particularly, it has been shown in \cite{nnbeam} that a well trained deep neural network can greatly reduce
  the time for power allocation with a satisfying performance loss compared to WMMSE approach. In addition, this is also a reason for some researchers to use reinforcement learning. For example, authors in \cite{bs1} use distributed Q learning that leads to a low-complexity sleep mode control algorithm for small cells.
  \emph{In summary, this motivation applies to literatures \cite{nnbeam,24,sun13,nnpower,auto-power,sun39,sun,75,knnbeam,U-Conf-1,bs1,S-3}}.
  \item \emph{Overcoming the Lack of Network Information/Knowledge:} Although centralized optimization approaches can achieve superior performance,
  they often needs to know global network information, which can be difficult to acquire. For example, the baseline scheme for BS switching in \cite{bs3}
  requires a full knowledge of traffic loads in prior, which is challenging to be precisely known in advance. However, with transfer learning,
  the past experience in BS switching can be utilized to guide current switching control even without the knowledge of traffic loads.
  To adjust handover parameters, fuzzy logic controller based approaches can be used. The controller is based on a set of pre-defined rules,
  each of which specifies a deterministic action under a certain system state. However, the setting of the action is highly dependent on expert knowledge about the network
  that
  can be unavailable for a new communication system or environment. In addition, knowing content popularity of users is the key to properly manage cache resource,
  and this popularity can be accurately learned by RNN and extreme learning machine. Moreover, model free reinforcement learning can help network nodes
  make optimized decisions without knowing the information about network dynamics. \emph{Overall, this motivation is a basic reason for adopting machine learning that applies to all the surveyed literatures.}
  \item \emph{Facilitating Self-organization Capabilities:} To reduce CAPEX and OPEX, and to simplify the coordination, optimization and configuration procedures of the network, self-organizing networks have been widely studied \cite{son1} and some researchers consider machine learning techniques as potential enablers to realize self-organization capabilities. By involving machine learning, especially reinforcement learning, each BS can self-optimize its resource allocation,
handover parameter configuration, and so on. \emph{In summary, this motivation applies to \cite{24,Power+,Add-1,sun40,sun43,3,vrchen,Conf-36,75,Backhaul+,Conf-30,Conf-34,benecache,bs1,bs10,M-Conf-2,M-Conf-3,M-4}}.
\item \emph{Reducing Signalling Overhead:} When distributed reinforcement learning is used, each learning agent only needs to
acquire partial network information to make a decision, which helps avoid large signalling overhead. On the contrary, traditional approaches
may require many information exchanges and hence lead to huge signalling cost. For example, as pointed out in \cite{S-21},
ad hoc on-demand distance vector routing will cause the constant flooding of routing messages in a CRN,
and the centralized approach taken as the baseline in \cite{sun39} allocates spectrum resource
based on the complete information about SUs.
\emph{This motivation has been highlighted in \cite{ha,sun13,sun39,sun40,75,Backhaul+,46,Conf-30,weicache,bs12,S-21,S-20,S-29}.}

  \item  \emph{Avoiding Past Faults:} For some heuristic and classical approaches that are based on fixed rules, they are incapable of avoiding unsatisfying results that have occurred previously, which means they are incapable of learning.
  Such approaches include the OSPF routing strategy taken as the baseline in \cite{S-29}, the handover strategy based on the comparison of RSSI values that is taken as the baseline by \cite{1}, the heuristic BS switching control strategies for comparison in \cite{tangjian}, max-SINR based user association, and so on.
  In \cite{S-29}, authors present an intuitive example, where OSPF routing leads to congestion at a router under a certain situation.
  Then, when this situation recurs, the OSPF routing protocol will make the same routing decision that causes congestion again.
  However, with deep learning being trained by using history network data,
  it can be predicted that whether a routing strategy will lead to congestion under the current traffic pattern.
  Other approaches listed face the same kind of problems. This issue can be overcome by reinforcement learning approaches in \cite{1,U-1,tangjian}, which evaluate
  each action based on its past performance. Hence, actions with bad performance can be avoided in the future.
  \emph{For surveyed literatures, this motivation applies to \cite{1,24,42,Add-1,vrchen,Conf-36,Backhaul+,46,benecache,weicache,drlcache,chenxian,U-1,tangjian,S-31,S-29,M-Conf-3}. }
  \item \emph{Learning Robust Patterns:} With the help of neural networks, useful patterns related to networks and users can be extracted.
  These patterns are useful in resource management, localization, and so on.
  Specifically, authors in \cite{nnpower} use a CNN to learn the spatial features of the channel gain matrix to make wiser power control decisions than WMMSE.
  For fingerprints based localization, traditional approaches, such as Horus, directly relies on the received signal strength data
  that can be easily affected by the complex indoor propagation environment. This fact has motivated researchers
  to improve localization accuracy by learning more robust fingerprint patterns using neural networks.
  \emph{For surveyed literatures, this motivation applies to \cite{nnbeam,1,3d,nnpower,auto-power,3,vrchen,S-1,drlcache,elmcache,chenjsac,chencran,tonycache,chenxian,waliduav,
  tangjian,S-20,S-29,yufei,M-Arxiv-1,ind16,ind31,ind32,ind4,xvyu,xvyu1}.}
  \item \emph{Achieving Better Performance than Traditional Optimization:} Traditional optimization methods include submodular optimization theory,
  dual decomposition, and so on. In \cite{knnbeam}, authors have demonstrated that their designed KNN based beam allocation algorithm can outperform
  a beam allocation algorithm based on submodular optimization theory with the increase of training data size, while it
  is shown in \cite{U-1} that reinforcement learning based user association achieves better performance than the approach based on dual decomposition.
  Hence, it can be inferred that machine learning has the potential in reaching better system performance compared to traditional optimization approaches.
  \emph{In surveyed literatures, this motivation applies to \cite{nnpower,knnbeam,U-1}.}
\end{itemize}

\section{Challenges and Open Issues}

Although many studies have been conducted on the applications of ML in wireless communications,
several challenges and open issues are
identified in this section to facilitate further research in this area.

\subsection{Machine Learning Based Heterogenous Backhaul/Fronthaul Management}

In future wireless networks, various backhaul/fronthaul solutions will coexist \cite{yan}, including wired backhaul/fronthaul like fiber and cable
as well as wireless backhaul/fronthaul like the sub-6 GHz band. Each solution has a different amount of energy consumption and different bandwidth,
and hence the management
of backhaul/fronthaul is important to the whole system performance. In this case, ML
based techniques can be utilized to select suitable backhaul/fronthaul solutions based on the extracted traffic patterns and performance requirements
of users.

%

\subsection{Infrastructure Update}

To make preparations for the deployment
of ML based communication systems, current wireless network infrastructures should be evolved.
For example, servers equipped with GPUs can be deployed at the network edge to implement deep learning based
signal processing, resource management and localization, and storage devices are needed at the network edge as well to achieve in-time data analysis.
Moreover, network function virtualization (NFV) should be involved in the wireless network, which decouples the network functions and hardware, and then
network functions can be implemented as softwares. On the basis of NFV, machine learning can be adopted to realize flexible network control and configuration.

\subsection{Machine Learning Based Network Slicing}


As a cost-efficient way to support diverse use cases, network slicing has been advocated by both academia and industry \cite{xiangslice}.
The core of network slicing
is to allocate appropriate resources including computing, caching, backhaul/fronthaul and radio resources on demand to guarantee the performance
requirements of different slices under slice isolation constraints.
Generally speaking, network slicing can benefit from ML
in the following aspects. First, ML can be used
to learn the mapping from service demands to resource allocation plans,
and hence a new network slice can be quickly constructed.
Second,
by employing transfer learning, knowledge about resource allocation plans for different use cases
in one environment can act as useful knowledge in another environment, which can speed up the learning process.
Recently, authors in \cite{slice1} and \cite{slice2} have applied DRL to network slicing,
and the advantages of DRL are demonstrated via simulations.

\subsection{Standard Datasets and Environments for Research}

To make researchers pay full attention to the learning algorithm design and conduct fair comparisons between different ML based approaches,
it is essential to identify some common problems in wireless networks together with corresponding labeled/unlabeled data for supervised/unsupervised learning based approaches, similar to the open dataset MNIST which is often used in computer vision.
For reinforcement learning based approaches, standard network control problems together with well defined environments should be built,
similar to the standard environment MountainCar-v0.

\subsection{Theoretical Guidance for Algorithm Implementation}

It is known that the performance of ML algorithms is affected by the selection of hyperparameters like learning rate, loss functions, and so on.
Trying different hyperparameters directly is a time-consuming task, especially when the training time for the model under a fixed set of hyperparameters is
long. Moreover, the theoretical analysis of the dataset size needed for training,
the performance bound of deep learning architectures, and the ability of generalization of different learning models are still open questions.
Since stability is one of the main features of communication systems, rigorous theoretical studies are essential to ensure ML based
approaches always work well in practical systems.

\subsection{Transfer Learning Based Approaches}

Transfer learning promises transferring the knowledge learned from one task to another similar task. By avoiding training
learning models from scratch, the learning process in new environments can be speeded up,
and the ML algorithm can have a good performance even with a small amount of training data.
Therefore, transfer learning is critical for the practical implementation of learning models considering the cost
for training without prior knowledge. Using transfer learning, network operators can
solve new but similar problems in a cost-efficient manner.
However, negative effects of prior knowledge on system performance
should be addressed as pointed out in \cite{bs3}, and need further investigation.

\section{Conclusions}

This paper surveys the state-of-the-art applications of ML in wireless communication and outlines several unresolved problems.
Faced with the intricacies of these applications, we have broadly divided the body of knowledge into resource management in the MAC layer, networking and mobility management in the network layer, and localization in the application layer.
Within each of these topics, we have surveyed the diverse ML based approaches that have been proposed for enabling wireless networks to run intelligently.
Nevertheless, considering that the applications of ML in wireless communications are still at the initial stage,
there are quite a number of problems that need further investigation. For example, infrastructure update is required for the implementation of ML based paradigms,
theoretical analysis on the ML based approaches should be conducted to provide a performance guarantee, and open data sets and environments are expected to facilitate
future research on the ML applications in a wide range.

\end{document}